\newif\ifCM 
\newif\ifSF 
\CMfalse
\SFfalse
\ifCM
\documentclass[final, 3p]{elsarticle}
\else
\documentclass[11pt,a4paper]{article}
\fi
\usepackage{float}
\ifCM
\usepackage[caption=false]{subfig}
\else
\usepackage{subfig}
\usepackage[ocgcolorlinks,allcolors={blue}]{hyperref}
\usepackage[numbers,sort&compress]{natbib}
\fi
\usepackage[utf8]{inputenc}
\usepackage[T1]{fontenc}
\usepackage{amsmath}
\usepackage{amssymb}
\usepackage{graphicx}
\usepackage[usenames,dvipsnames,svgnames,table]{xcolor}
\usepackage{multirow}
\usepackage{xspace}
\usepackage{ifthen}
\usepackage{ifmtarg}
\usepackage{calc}
\usepackage{algpseudocode}
\usepackage{algorithm}
\usepackage{algorithmicx}
\usepackage{lineno}
\usepackage{placeins}
\usepackage{appendix}
\ifCM
\else
\fi
\ifSF
\else
\graphicspath{{Images/}}
\fi
\makeatletter
\DeclareFontFamily{U}{mathx}{\hyphenchar\font45}
\DeclareFontShape{U}{mathx}{m}{n}{
      <5> <6> <7> <8> <9> <10>
      <10.95> <12> <14.4> <17.28> <20.74> <24.88>
      mathx10
      }{}
\DeclareSymbolFont{mathx}{U}{mathx}{m}{n}
\DeclareFontSubstitution{U}{mathx}{m}{n}
\DeclareMathSymbol{\bigtimes}{1}{mathx}{"91}
\newcommand*{\ds}{\displaystyle}
\newcommand*{\Reel}{{\mathbb{R}}}      
\newcommand*{\Poly}{{\mathbb{P}}}      
\newenvironment{Proof}{\noindent \textbf{Proof}.}{
\begin{flushright}
$\square$
\end{flushright}
}
\newcommand*{\CA}{\texttt{code\_aster }} 
\newcommand*{\psv}[3][]{
	\@ifmtarg{#1}{
                (#2,#3) 
        }{%
                (#2,#3)_{\vecteur{L}^2(#1)}
        }%
} 
\newcommand*{\psm}[3][]{
	\@ifmtarg{#1}{
                (#2,#3) 
        }{%
                (#2,#3)_{\matrice{L}^2(#1)}
        }%
}%
\newcommand*{\psmd}[3][]{
	\@ifmtarg{#1}{
                (#2,#3) 
        }{%
                (#2,#3)_{\matrice{L}^2_{Q}(#1)}
        }%
} 
\newcommand*{\psd}[3][]{
	\@ifmtarg{#1}{
                (#2,#3) 
        }{%
                (#2,#3)_{L^2_{Q}(#1)}
        }%
}
\newcommand*{\psvd}[3][]{
	\@ifmtarg{#1}{
                (#2,#3) 
        }{%
                (#2,#3)_{\vecteur{L}^2_{Q}(#1)}
        }%
} 
\newcommand*{\ps}[3][]{
	\@ifmtarg{#1}{
                (#2,#3) 
        }{%
                (#2,#3)_{#1}
        }%
}
\newcommand*{\normemd}[2][]{
	\@ifmtarg{#1}{
                \Vert #2 \Vert 
        }{%
                \Vert #2 \Vert_{\matrice{L}^2_{Q}(#1)}
        }%
} 
\newcommand*{\normev}[2][]{
	\@ifmtarg{#1}{
                \Vert #2 \Vert 
        }{%
                \Vert #2 \Vert_{\vecteur{L}^2(#1)}
        }%
}
\newcommand*{\normem}[2][]{
	\@ifmtarg{#1}{
                \Vert #2 \Vert 
        }{%
                \Vert #2 \Vert_{\matrice{L}^2(#1)}
        }%
}
\newcommand*{\norme}[2][]{
	\@ifmtarg{#1}{
                \Vert #2 \Vert 
        }{%
                \Vert #2 \Vert_{#1}
        }%
}
\newcommand{\SCAL}{{\cdot}}
\newcommand*{\snorme}[2][]{
	\@ifmtarg{#1}{
                \vert #2 \vert 
        }{%
                \vert #2 \vert_{#1}
        }%
} 
\newcommand*{\tenseur}[3][\boldsymbol]{%
    \ifthenelse{#2=0}{#3}{%
	\ifthenelse{#2=1}{%
						\underline{#1{#3}}
					}{
		\ifthenelse{#2=2}{%
							\underline{\underline{ #1{#3}}}
						}{
			\ifthenelse{#2=3}{%
								\underline{\underline{\underline{#1{#3}}}}
							}{
				\ifthenelse{#2=4}{%
				{#1{\mathbb{#3}}}
				}{ERROR}
			}
		}
	}
}}
\newcommand*{\vecteur}[1]{\tenseur{1}{#1}} 
\newcommand*{\matrice}[1]{\tenseur{2}{#1}} 
\newcommand*{\divergence}[1]{\nabla {\cdot} #1 } 

\newcommand{\grads}{\matrice{\nabla}^{s}}
\newcommand{\dev}{\textrm{dev}}
\newcommand{\trace}{\textrm{trace}}
\newcommand{\Msym}{\Reel^{d\times d}_{\textrm{sym}}} 
\newcommand{\loadext}{\vecteur{f}} 
\newcommand{\Tn}{\vecteur{t}} 

\newcommand*{\Bn}{\Gamma_{\textrm{N}}}
\newcommand*{\Bd}{\Gamma_{\textrm{D}}}
\newcommand{\dT}{{\partial T}}
\newcommand{\FT}{\mathcal{F}_{\dT}} 
\newcommand{\Fh}{\mathcal{F}_h} 
\newcommand{\Th}{\mathcal{T}_h} 
\newcommand{\Fhi}{\mathcal{F}_h^{\textrm{i}}} 
\newcommand{\Fhb}{\mathcal{F}_h^{\textrm{b}}} 
\newcommand{\Fhbd}{\mathcal{F}_h^{\textrm{b,D}}} 
\newcommand{\Fhbn}{\mathcal{F}_h^{\textrm{b,N}}} 
\newcommand{\nTp}{\vecteur{n}{}_{T_+}} 
\newcommand{\nTm}{\vecteur{n}{}_{T_-}} 
\newcommand{\nT}{\vecteur{n}{}_{T}} 
\newcommand{\Pkd}{\Poly_d^k} 
\newcommand{\Pld}{\Poly_d^l} 
\newcommand{\Pkpd}{\Poly_d^{k+1}} 
\newcommand{\PkF}{\Poly_{d-1}^k} 
\newcommand{\UkT}{\vecteur{U}^k_T} 
\newcommand{\Ukhz}{\vecteur{U}^k_{h,0}} 
\newcommand{\Ukh}{\vecteur{U}^k_h} 
\newcommand{\Uknhd}{\vecteur{U}^{k,n}_{h,\textrm{D}}} 
\newcommand{\Ukpnhd}{\vecteur{U}^{k,n-1}_{h,\textrm{D}}} 
\newcommand{\UklT}{\vecteur{U}^{k,l}_T} 
\newcommand{\IVGTh}{\tilde{\vecteur{\mathcal{X}}}{\vphantom{\vecteur{\mathcal{X}}}}_{\Th}^{k_Q}} 
\newcommand{\vT}{\vecteur{v}_T} 
\newcommand{\vF}{\vecteur{v}_F} 
\newcommand{\vdT}{\vecteur{v}_{\partial T}} 
\newcommand{\vTh}{\vecteur{v}_{\Th}} 
\newcommand{\vFh}{\vecteur{v}_{\Fh}} 
\newcommand{\uT}{\vecteur{u}_T} 
\newcommand{\udT}{\vecteur{u}_{\dT}} 
\newcommand{\uTh}{\vecteur{u}_{\Th}} 
\newcommand{\uFh}{\vecteur{u}_{\Fh}} 

\newcommand{\uTn}{\vecteur{u}_T^n} 
\newcommand{\uTpn}{\vecteur{u}_{T_+}^n} 
\newcommand{\uTmn}{\vecteur{u}_{T_-}^n} 
\newcommand{\udTn}{\vecteur{u}_{\dT}^n} 
\newcommand{\udTpn}{\vecteur{u}_{\dT_+}^n} 
\newcommand{\udTmn}{\vecteur{u}_{\dT_-}^n} 
\newcommand{\uThn}{\vecteur{u}_{\Th}^n} 
\newcommand{\uFhn}{\vecteur{u}_{\Fh}^n} 

\newcommand{\vh}{\vecteur{v}_h}

\newcommand{\duT}{ \delta\vecteur{u}_T}

\newcommand{\dvT}{\delta \vecteur{v}_T}
\newcommand{\dudT}{ \delta\vecteur{u}_{\dT}}

\newcommand{\dvdT}{\delta \vecteur{v}_{\dT}}
\newcommand{\dvF}{\delta \vecteur{v}_F}
\newcommand{\duTh}{ \delta\vecteur{u}_{\Th}}

\newcommand{\dvTh}{\delta \vecteur{v}_{\Th}}
\newcommand{\duFh}{ \delta\vecteur{u}_{\Fh}}

\newcommand{\dvFh}{\delta \vecteur{v}_{\Fh}}
\newcommand{\vu}{\vecteur{u}}
\newcommand{\vv}{\vecteur{v}}

\newcommand{\EkT}{\matrice{E}^k_T} 
\newcommand{\EkTp}{\matrice{E}^k_{T_+}} 
\newcommand{\EkTm}{\matrice{E}^k_{T_-}} 
\newcommand{\SdTk}{\vecteur{S}_{\dT}^k}
\newcommand{\SdTpk}{\vecteur{S}_{\dT_+}^k}
\newcommand{\SdTmk}{\vecteur{S}_{\dT_-}^k}

\newcommand{\SdTks}{\vecteur{S}_{\dT}^{k*}}
\newcommand{\SdTpks}{\vecteur{S}_{\dT_+}^{k*}}
\newcommand{\SdTmks}{\vecteur{S}_{\dT_-}^{k*}}
\newcommand{\DTkp}{\vecteur{D}_T^{k+1}}
\newcommand{\PikT}{\matrice{\Pi}^{k}_T} 
\newcommand{\PikF}{\matrice{\Pi}^{k}_F} 
\newcommand{\PikdT}{\matrice{\Pi}^{k}_{\partial T}} 
\newcommand{\strain}{\matrice{\varepsilon}} 
\newcommand{\estrain}{\matrice{\varepsilon}^e} 
\newcommand{\pstrain}{\matrice{\varepsilon}^p} 
\newcommand{\IF}{\vecteur{q}} 
\newcommand{\IV}{\vecteur{\alpha}} 
\newcommand{\IVG}{\vecteur{\chi}} 
\newcommand{\IVGT}{\tilde{\vecteur{\chi}}_T} 
\newcommand{\IVGh}{\tilde{\vecteur{\chi}}_{\Th}} 
\newcommand{\stress}{\matrice{\sigma}} 
\newcommand{\elasticmodule}{{\tenseur{4}{\mathbb{C}}}} 
\newcommand{\epmodule}{{\tenseur{4}{\mathbb{C}}{}_{ep}}} 
\newcommand{\depmodule}{{\tilde{\tenseur{4}{\mathbb{C}}}{}_{ep}}} 
\newcommand{\depmoduleni}{{\tilde{\tenseur{4}{\mathbb{C}}}{}_{ep}^{n,i}}} 

\newcommand{\epmodulenew}{{\tenseur{4}{\mathbb{C}}{}_{ep}^{\textrm{new}}}} 
\newcommand{\epmodulen}{{\tenseur{4}{\mathbb{C}}{}_{ep}^{n}}} 
\newcommand{\depmodulen}{{\tilde{\tenseur{4}{\mathbb{C}}}{}_{ep}^{n}}} 
\newcommand{\Qp}{\vecteur{\xi}} 
\newcommand{\Wp}{\omega}

\newcommand{\PikTmd}{\matrice{\Pi}^{k}_{Q,T}} 
\newcommand{\PikTpmd}{\matrice{\Pi}^{k}_{Q,T_+}} 
\newcommand{\PikTmmd}{\matrice{\Pi}^{k}_{Q,T_-}} 
\newcommand{\dstress}{\tilde{\stress}} 

\newcommand{\mm}{\textrm{mm}}

\newcommand{\MPa}{\textrm{MPa}}
\newcommand{\GPa}{\textrm{GPa}}
\ifCM
\newtheorem{theorem}{Theorem}
\newtheorem{lemma}[theorem]{Lemma}
\newtheorem{hypothesis}[theorem]{Hypothesis}
\newtheorem{proposition}[theorem]{Proposition}
\newtheorem{corollary}[theorem]{Corollary}
\newtheorem{definition}[theorem]{Definition}
\newtheorem{remark}[theorem]{Remark}
\newtheorem{prop}[theorem]{Proposition}
\else
\newtheorem{theorem}{Theorem}
\newtheorem{lemma}[theorem]{Lemma}
\newtheorem{hypothesis}[theorem]{Hypothesis}

\newtheorem{remark}[theorem]{Remark}

\fi
\makeatother
\ifCM
\journal{Computer Methods in Applied Mechanics and Engineering}
\biboptions{sort&compress}
\fi
\begin{document}
\ifCM
%
%

%

%
\begin{abstract}
We devise and evaluate numerically a Hybrid High-Order (HHO) method for incremental associative plasticity with small deformations. The HHO method uses as discrete unknowns piecewise polynomials of order $k\ge1$ on the mesh skeleton, together with cell-based polynomials that can be eliminated locally by static condensation.  The HHO method supports polyhedral meshes with non-matching interfaces, is free of volumetric locking, and the integration of the behavior law is performed only at cell-based quadrature nodes. Moreover,  the principle of virtual work is satisfied locally with equilibrated tractions. Various two- and three-dimensional test cases from the literature are presented including comparison against known solutions and against
 results obtained with an industrial software using conforming and mixed finite elements.
\end{abstract}

\end{frontmatter}
\else
\author{Mickaël Abbas$^1$,  Alexandre Ern$^{2,3}$ and Nicolas Pignet$^{1,2,3}$}
\title{A Hybrid High-Order method for incremental associative plasticity with small deformations}
\maketitle
\footnotetext[1]{EDF R\&D, 7 Boulevard Gaspard Monge, 91120 Palaiseau,
  France and IMSIA, UMR EDF/CNRS/CEA/ENSTA 9219, 828 Boulevard des Maréchaux,
91762 Palaiseau Cedex, France}
\footnotetext[2]{Université Paris-Est, CERMICS (ENPC), 6-8 avenue Blaise Pascal, 77455 Marne-la-Vall\'ee cedex 2, France}
\footnotetext[3]{INRIA, 75589 Paris, France}
\begin{abstract}
We devise and evaluate numerically a Hybrid High-Order (HHO) method for incremental associative plasticity with small deformations. The HHO method uses as discrete unknowns piecewise polynomials of order $k\ge1$ on the mesh skeleton, together with cell-based polynomials that can be eliminated locally by static condensation.  The HHO method supports polyhedral meshes with non-matching interfaces, is free of volumetric locking, and the integration of the behavior law is performed only at cell-based quadrature nodes. Moreover,  the principle of virtual work is satisfied locally with equilibrated tractions. Various two- and three-dimensional test cases from the literature are presented including comparison against known solutions and against
 results obtained with an industrial software using conforming and mixed finite elements.
\end{abstract}
\\
\fi
%
%
\section{Introduction}
Hybrid High-Order (HHO) methods have been introduced a few years ago for diffusion problems in \cite{DiPEL:2014} and for linear elasticity problems in \cite{DiPEr:2015}. Recently, the development of HHO methods has received a vigorous interest. Examples include  in solids mechanics Biot's problem \cite{BoBoD:2016}, nonlinear elasticity with small deformations \cite{BoDPS:2017}, and hyperelasticity with finite deformations \cite{AbErPi2018}, and in fluid mechanics, the incompressible Stokes equations \cite{DPELS:16}, the steady incompressible Navier--Stokes equations \cite{DiPKr:2017}, and viscoplastic flows with yield stress \cite{Cascavita2018}.
The discrete unknowns in HHO methods are face-based unknowns that are piecewise polynomials on the mesh skeleton. Cell-based unknowns are also introduced. These additional unknowns are instrumental for the stability and approximation properties of the method and can be locally eliminated by using the well-known static condensation technique (based on a local Schur complement). For nonlinear problems, this elimination is performed at each step of the nonlinear iterative solver  (typically Newton's method). 

The devising of HHO methods hinges on two key ideas: \textup{(i)} a local higher-order reconstruction operator acting on the face and cell unknowns; \textup{(ii)} a local stabilization operator that weakly enforces on each mesh face the consistency between the local face unknowns and the trace of the cell unknowns. A somewhat subtle design of the stabilization operator has been proposed in \cite{DiPEr:2015,DiPEL:2014} leading to $O(h^{k+1})$ energy-error estimates for linear model problems with smooth solutions, where $h$ is the mesh-size and $k$ is the polynomial order of the face unknowns. 
HHO methods offer several advantages: \textup{(i)} the construction is dimension-independent; \textup{(ii)} general meshes (including fairly general polyhedral mesh cells and non-matching interfaces) are supported; \textup{(iii)} a local formulation using equilibrated fluxes is available, and \textup{(iv)}  computational benefits owing to the static condensation of the cell unknowns and the higher-order convergence rates. In computational mechanics, another salient feature of HHO methods is the absence of volumetric locking \cite{DiPEr:2015}. Furthermore, HHO methods have been bridged in \cite{CoDPE:2016} to Hybridizable Discontinuous Galerkin (HDG) methods \cite{CoGoL:09} and to nonconforming Virtual Element Methods (ncVEM) \cite{Ayuso2016}. The essential difference with HDG methods is 
that the HHO stabilization is different so as to deliver higher-order convergence rates on general meshes. Concerning ncVEM, the devising viewpoint is different (ncVEM considers the computable projection of virtual functions instead of a reconstruction operator), and the stabilization achieves similar convergence rates as HHO but is written differently. An open-source implementation of HHO methods, the \texttt{DiSk++} library, is available using generic programming tools \cite{CicDE:2017Submitted}.

In the present work, we devise and evaluate numerically a HHO method for incremental associative plasticity with small deformations. Modelling plasticity problems is particularly relevant in nonlinear solid mechanics since this is one of the main nonlinearities that can be encountered. Moreover, plasticity can have a major influence on the behavior of a mechanical structure. Plastic deformations are generally assumed to be incompressible, which can lead to serious volumetric-locking problems, particularly with a continuous Galerkin (cG) approximation based on (low-order) $H^1$-conforming finite elements, where only the displacement field is approximated globally contrary to the plastic deformations and the variables associated with the plastic behavior which are defined and solved locally in each mesh cell.
A way to circumvent volumetric locking is to consider mixed methods on simplicial or hexahedral meshes, as in \cite{Chiumenti2004, Simo1985,Simo1990}. However, mixed methods need additional global unknowns to impose the condition of plastic incompressibility that generally increase the cost of building and solving the global system (the variables associated with the plastic behavior are still solved locally). Moreover, devising mixed methods on polyhedral meshes with non-matching interfaces is a delicate question. Note that cG methods as well as mixed methods require to perform the integration of the behavior law only at quadrature nodes in the mesh cells. Another class of methods free of volumetric locking are discontinuous Galerkin (dG) methods. We mention in particular \cite{Eyck2006, Eyck2008, Eyck2008a, Noels2006} for hyperelasticity and \cite{Molari2006} for damage mechanics. Interior penalty dG methods have been developed for classical plasticity with small \cite{Hansbo2010, Liu2010} and finite \cite{Liu2013} deformations, and for gradient plasticity with small  \cite{Djoko2007, Djoko2007a} and finite \cite{McBride2009} deformations. However, dG methods from the literature generally require to perform the integration of the behavior law also at additional quadrature nodes located at the mesh faces. Since the behavior integration can be the most expensive part of the computation during the assembling step \cite{Mfront}, this additional integration on the mesh faces can lead to a substantial increase in the computational burden. Moreover, implementing the behavior integration on the mesh faces in an existing finite element code is not straightforward since the data structure for internal variables cannot be necessarily re-used. We also mention the lowest-order Virtual Element Method (VEM) for inelastic problems with small deformations devised in~\cite{BeiraodaVeiga2015} using the maximum norm of the tangent modulus for stabilization (see also \cite{Artioli2017} for a two-dimensional higher-order extension), whereas the case of finite deformations is treated in~\cite{Wriggers2017a}, still in the lowest-order case, with a pseudo-energy based stabilization.

The HHO method devised in this work uses polynomials of arbitrary order $k\ge1$ on the mesh faces, and as in \cite{BoDPS:2017,AbErPi2018}, the local reconstruction operator builds a symmetric gradient in the tensor-valued polynomial space $\Pkd(T; \Msym)$, where $T$ is a generic mesh cell and $d$ is the space dimension. The present HHO method offers the above-mentioned benefits of HHO methods (dimension-independent construction, general meshes, local conservation, static condensation). In particular, equilibrated tractions satisfying the law of action and reaction while being in local balance with the external loads and the internal efforts are available. Moreover, in view of the above discussion on the literature, we observe that the present HHO method \textup{(i)} hinges on a displacement-based formulation thus avoiding the need to introduce additional global unknowns; \textup{(ii)} is free of volumetric locking; \textup{(iii)} supports general meshes; and \textup{(iv)} requires the integration of the behavior law only at the cell level. Another attractive feature is that the linear system at each step of Newton's method is coercive for strain-hardening materials provided the stabilization parameter is simply positive. We also notice that, to our knowledge, HDG methods have not yet been devised for plasticity problems (for hyperelasticity problems, we mention \cite{KaLeC:2015, Nguyen2012a}). Owing to the close links between HHO and HDG methods, this work can thus be seen as the first HDG-like method for plasticity problems. Another follow-up of the HHO idea of local reconstruction is to pave the way for dG methods requiring only a cell-based integration of the behavior law, by using discrete gradients as in \cite{Eyck2006,DiPietro2011,KaLeC:2015}. 
We also mention the recent study of low-order hybrid dG method with conforming traces and the hybridizable weakly conforming Galerkin method with nonconforming traces in \cite{Bayat2018} in the context of nonlinear solid mechanics. 

This paper is organized as follows: in Section~2, we present the incremental associative plasticity problem and the weak formulation of the governing equations. In Section~3, we devise the HHO method and highlight some of its theoretical aspects. In Section~4, we investigate numerically the HHO method on two- and three-dimensional test cases from the literature, and we compare our results to analytical solutions whenever available and to numerical results obtained using established cG and mixed methods implemented in the open-source industrial software \CA \cite{CodeAster}. 
\section{Plasticity model}\label{sec::model}
In what follows, we write $v$ for scalar-valued fields, $\vv$ or $\vecteur{V}$ for vector-valued fields, $\matrice{V}$ for second-order tensor-valued fields, and $\tenseur{4}{V}$ for fourth-order tensor-valued fields. Contrary to the elastic model, the elastoplastic model is based on the assumption that the deformations are no longer reversible. We place ourselves within the framework of generalized standard materials initially introduced in  \cite{Halphen1975} and further developed in \cite{Lemaitre1994}. Moreover, we consider the regime of small deformations and the plasticity model is assumed to be strain-hardening (or perfect) and rate-independent, i.e., the time and the speed of the deformations have no influence on the solution. For this reason, only the incremental plasticity problem is considered.
\subsection{Helmholtz free energy and yield function}\label{ss:helmhotz}
An important hypothesis for the modelling of plasticity under small deformations is that the (symmetric) total strain tensor $\strain$, which is equal to the symmetric gradient $\grads\vu$ of the displacement field $\vu$, can be decomposed into the sum of an elastic part and a plastic part, denoted by $\estrain$ and $\pstrain$ respectively, so that $\strain = \estrain + \pstrain $, which is rewritten as follows:
\begin{equation}
\estrain = \strain - \pstrain.
\end{equation}
Both tensors, $\estrain$ and $\pstrain$, are symmetric, and the plastic deformations are assumed to be incompressible so that
\begin{equation}
\trace(\pstrain) =0.
\end{equation}
Since we consider generalized standard materials, the local material state is described by the total strain tensor $\strain\in\Msym$, the plastic strain tensor $\pstrain\in\Msym$, and a finite collection of internal variables $\IV:= ( \alpha_1, \cdots, \alpha_m) \in \Reel^m$. We assume that there exists a Helmholtz free energy $\Psi: \Msym \times \Reel^m \rightarrow \Reel$ acting on the pair $(\estrain, \IV)$ and satisfying the following hypothesis.
\begin{hypothesis}[Helmholtz free energy]\label{hyp_free_ener}
$\Psi$ can be decomposed additively into an elastic and a plastic part as follows:
\begin{equation}
 \Psi(\estrain, \IV) = \frac{1}{2}\, \estrain : \elasticmodule: \estrain + \Psi^p( \IV)
 \end{equation} 
where the function $\Psi^p$ is strictly convex, and the elastic modulus is $\elasticmodule = 2\mu \tenseur{4}{I}^{s} + \lambda \matrice{I} \otimes \matrice{I}$, with $\mu>0$, $3\lambda + 2 \mu > 0$, $(\tenseur{4}{I}^{s})_{ij,kl}= \frac{1}{2}(\delta_{ik}\delta_{jl} + \delta_{il}\delta_{jk})$, and $(\matrice{I} \otimes \matrice{I})_{ij,kl}=\delta_{ij}\delta_{kl}$. 
\end{hypothesis}
The elastic modulus $\elasticmodule$ is isotropic, constant, and positive definite with $\matrice{e} : \elasticmodule : \matrice{e} = 2 \mu \, \matrice{e}: \matrice{e} +\lambda\trace(\matrice{e})^2$, for all $\matrice{e} \in \Msym$.

As a consequence of the second principle of thermodynamics, the Cauchy stress tensor $\stress \in \Msym$ and the thermodynamic forces $\IF \in \Reel^{m}$  are derived from $\Psi$ as follows:
\begin{align}
\stress = \partial_{ \estrain} \Psi = \elasticmodule : \estrain \quad \textrm{ and } \quad
\IF =   \partial_{ \IV} \Psi^p. \label{eq_IF}
\end{align}
The criterion to determine whether the deformations are plastic hinges on the scalar yield function $\Phi: \Msym \times \Reel^m \rightarrow \Reel$, which is a continuous and convex function of the stress tensor $\stress$ and the thermodynamic forces $\IF$. The set of stresses and thermodynamic forces that verify $\Phi(\stress, \IF) \leq 0$ is the convex set of admissible states (or plasticity admissible domain):
\begin{equation}
\mathcal{A} := \left\lbrace (\stress, \IF) \in \Msym \times \Reel^m \: | \: \Phi(\stress, \IF) \leq 0 \right\rbrace.
\end{equation}
The set of admissible states is partitioned into two disjoint subsets, the elastic domain $\mathcal{A}^e$ which is the interior of the set $\mathcal{A}$ and the yield surface $\partial \mathcal{A}$ which is the boundary of the set $\mathcal{A}$, so that
\begin{equation}
\mathcal{A}^e := \left\lbrace (\stress, \IF) \in\mathcal{A} \: | \: \Phi(\stress, \IF) < 0 \right\rbrace, \quad \partial\mathcal{A} := \left\lbrace (\stress, \IF) \in \mathcal{A} \: | \: \Phi(\stress, \IF) = 0 \right\rbrace.
\end{equation}
\begin{hypothesis}[Yield function]\label{hyp_yield}
The yield function $\Phi: \Msym \times \Reel^m \rightarrow \Reel$ satisfies the following properties: \textup{(i)} $\Phi$ is a piecewise analytical function; \textup{(ii)} the point $(\matrice{0}, \vecteur{0})$  lies in the elastic domain, i.e., $\Phi(\matrice{0}, \vecteur{0}) < 0$; and \textup{(iii)} $\Phi$ is differentiable at all points on the yield surface $\partial\mathcal{A}$.
\end{hypothesis}
\subsection{Plasticity model problem in incremental form}\label{ss:mech_model}
We are interested in finding the quasi-static evolution in the pseudo-time interval $[0,t_F]$, $t_F > 0$, of an elastoplastic material body that occupies the domain $\Omega_0$ in the reference configuration. Here, $\Omega_0 \subset \Reel^d$, $d \in \{2,3\}$, is a bounded connected polyhedral domain with  Lipschitz boundary $\Gamma := \partial \Omega$ decomposed in the two relatively open subsets $\Bn$ and $\Bd$, where a Neumann and a Dirichlet condition are enforced respectively, and such that $\overline{\Bn} \cup \overline{\Bd} = \Gamma$, $\Bn \cap \Bd = \emptyset $, and $\Bd$ has positive Hausdorff-measure (so as to prevent rigid-body motions). The evolution occurs under the action of a body force $\loadext:\Omega_0 \times [0,t_F] \rightarrow \Reel^d $, a traction force $\Tn: \Bn \times [0,t_F] \rightarrow \Reel^d$ on the Neumann boundary $\Bn$, and a prescribed displacement $\vu{}_{\textrm{D}}: \Bd \times [0,t_F] \rightarrow \Reel^d$ on the Dirichlet boundary $\Bd$:
The pseudo-time interval $[0, t_F]$ is discretized into $N$ subintervals such that $t^0=0 < t^1 < \cdots < t^N = t_F$.
We denote by $V$, resp. $V_0$, the set of all kinematically admissible displacements which satisfy the Dirichlet conditions, resp. homogeneous Dirichlet conditions on $\Bd$ 
\ifCM
\begin{equation}
V_{\textrm{D}}^n = \left\lbrace \vv \in H^1(\Omega_0; \Reel^d) \: | \: \vv = \vu{}_{\textrm{D}}(t^n) \: \mbox{ on } \Bd \right\rbrace, \quad V_0 = \left\lbrace \vv \in H^1(\Omega_0; \Reel^d) \: | \: \vv = \vecteur{0} \: \mbox{ on } \Bd \right\rbrace.
\end{equation}
\else
\begin{equation}
 V_{\textrm{D}}^n = \left\lbrace \vv \in H^1(\Omega_0; \Reel^d) \: | \: \vv = \vu{}_{\textrm{D}}(t^n) \: \mbox{ on } \Bd \right\rbrace, \; V_0 = \left\lbrace \vv \in H^1(\Omega_0; \Reel^d) \: | \: \vv = \vecteur{0} \: \mbox{ on } \Bd \right\rbrace.
\end{equation}
\fi
Moreover, we denote $\IVG:=( \pstrain, \IV ) \in \vecteur{X}$ the generalized internal variables, where the space of the generalized internal variables is
\begin{equation} \label{eq:def_space_X}
\vecteur{X} := \left\lbrace \IVG=(\pstrain, \IV) \in  \Msym \times \Reel^{m}  \: | \: \trace(\pstrain)= 0 \right\rbrace.
\end{equation} 
Then the problem can be formulated as follows: For all $1 \leq n \leq N$, given $\vu^{n-1} \in V_{\textrm{D}}^{n-1}$ and $\IVG^{n-1} \in  L^2(\Omega_0; \vecteur{X})$ from the previous pseudo-time step or the initial condition,  find $\vu^n \in V_{\textrm{D}}^n$ and $\IVG^n \in  L^2(\Omega_0; \vecteur{X})$ such that
\begin{subequations}\label{weak_form_cont}
\begin{equation} 
  \int_{\Omega_0} \stress^n : \strain(\vv) \,d\Omega_0 = \int_{\Omega_0} \loadext^n \SCAL \vv \,d\Omega_0 + \int_{\Bn} \Tn^n \SCAL \vv \,d\Gamma  \textrm{ for all } \vv\in V_0,
\end{equation}
\textrm{and}
\begin{equation}
( \IVG^n, \stress^n, \epmodulen) =  \textrm{PLASTICITY}(\IVG^{n-1}, \strain^{n-1}, \strain^n-\strain^{n-1} ).
\end{equation}
\end{subequations}
The procedure $\textrm{PLASTICITY}$ allows one to compute the new values of the generalized internal variables $\IVG$, the stress tensor $\stress$ and the consistent elastoplastic tangent modulus $\epmodule$ at each pseudo-time step. This procedure is detailed in Section~\ref{ss:flow_rules} below. The incremental problem \eqref{weak_form_cont} can be reformulated as an incremental variational inequality by introducing a dissipative function, see for example \cite{Djoko2007a}. For strain-hardening plasticity, the weak formulation \eqref{weak_form_cont} is well-posed, see \cite[Section 6.4]{Han2013}. For perfect plasticity, under additional hypotheses on the loads, the existence of a solution to \eqref{weak_form_cont} with bounded deformation is studied in \cite{Maso2006}.
\subsection{Algorithmic aspects}\label{ss:flow_rules}
Algorithm \ref{algo::plasticity} presents the incremental associative elastoplasticity problem that has to be solved in order to find the new value, after incrementation,  of the generalized internal variables $\IVG^{\textrm{new}} = (\strain^{\textrm{p,new}}, \IV^{\textrm{new}}) \in \vecteur{X}$, the stress tensor $\stress^{\textrm{new}} \in \Msym$, and the consistent elastoplastic tangent modulus $\epmodulenew$, given the generalized internal variables $\IVG \in \vecteur{X}$, the strain tensor $\strain \in \Msym$, and the incremental strain tensor $\mathrm{d}{\strain} \in \Msym$. Solving this problem is denoted as previously
\begin{equation}
(\IVG^{\textrm{new}}, \stress^{\textrm{new}}, \epmodulenew)= \textrm{PLASTICITY}(\IVG, \strain, \mathrm{d}{\strain}).
\end{equation}
The procedure to compute $(\IVG^{\textrm{new}}, \stress^{\textrm{new}}, \epmodulenew)$ is described in Algorithm~\ref{algo::plasticity}.
First, an elastic trial state $(\stress^{\textrm{trial}}, \IF^{\textrm{trial}})$ is computed. If $(\stress^{\textrm{trial}}, \IF^{\textrm{trial}}) \in \mathcal{A}^e$, then the evolution is elastic, the trial state is accepted, and the internal variables are not modified. Otherwise, the evolution is plastic and the normal and flow rules are used to correct the elastic trial state  $(\stress^{\textrm{trial}}, \IF^{\textrm{trial}})$. Specifically, we introduce (see line~\ref{line_Lambda} of Algorithm~\ref{algo::plasticity}) the plastic multiplier (or consistency parameter) which depends on the yet-unknown pair $(\stress^{\textrm{new}}, \IF^{\textrm{new}})$,
\begin{equation}
\Lambda(\stress^{\textrm{new}}, \IF^{\textrm{new}}) = \dfrac{ \partial_{\stress} \Phi(\stress^{\textrm{new}}, \IF^{\textrm{new}}) : \elasticmodule : \mathrm{d}{\strain}}{\partial_{\stress} \Phi(\stress^{\textrm{new}}, \IF^{\textrm{new}}):\elasticmodule:\partial_{\stress} \Phi(\stress^{\textrm{new}}, \IF^{\textrm{new}}) + \overline{H}(\stress^{\textrm{new}}, \IF^{\textrm{new}})} \ge 0,
\end{equation}
where $\overline{H}(\stress^{\textrm{new}}, \IF^{\textrm{new}}) := \partial_{\IF} \Phi(\stress^{\textrm{new}}, \IF^{\textrm{new}}): \partial^2_{\IV, \IV} \Psi^p(\IV^{\textrm{new}}) : \partial_{\IF} \Phi(\stress^{\textrm{new}}, \IF^{\textrm{new}})\geq 0$ is the generalized hardening modulus. For strain-hardening plasticity (resp. perfect plasticity), we have $\overline{H}>0$ (resp. $\overline{H}=0$). The normal and flow rules then state that the pair $(\stress^{\textrm{new}}, \IF^{\textrm{new}}) \in \mathcal{A}$ must be such that
$\mathrm{d}{\pstrain} = \Lambda \, \partial_{\stress} \Phi$ and $\mathrm{d}{\IV} = - \Lambda \, \partial_{\IF} \Phi$, where the dependencies in $(\stress^{\textrm{new}}, \IF^{\textrm{new}})$ are omitted for simplicity and where the multiplier $\Lambda$ verifies in addition the complementary conditions
\begin{equation}\label{complementary_conditions}
\Lambda \geq 0, \quad  \Lambda \, \Phi = 0,
\end{equation} 
and the consistency condition
\begin{equation}\label{consitency_condition}
\Lambda \, \mathrm{d}{\Phi} = 0 \quad \text{if $(\stress^{\textrm{new}}, \IF^{\textrm{new}}) \in \partial\mathcal{A}$}.
\end{equation}
For strain-hardening plasticity, one can show (see \cite{Hill1958}) that there exists a unique solution to the constrained nonlinear system considered in lines~\ref{line_sept}-\ref{line_Lambda} of Algorithm \ref{algo::plasticity}. 
\begin{algorithm}[H]
\caption{Computation of $(\IVG^{\textrm{new}}, \stress^{\textrm{new}}, \epmodulenew)$}\label{algo::plasticity}
\begin{algorithmic}[1] 
        \Procedure{Plasticity}{$\IVG, \strain, \mathrm{d}\strain$}
            \State Set $\stress^{\textrm{trial}} = \elasticmodule : (\estrain + \mathrm{d}\strain)$ and $\IF^{\textrm{trial}} = \partial_{\IV}\Psi^p(\IV)$
             \If{ $(\stress^{\textrm{trial}}, \IF^{\textrm{trial}}) \in \mathcal{A}^e$ \textbf{or if}
           $(\stress^{\textrm{trial}}, \IF^{\textrm{trial}}) \in \partial\mathcal{A}$ and $\partial_{\stress}\Phi(\stress^{\textrm{trial}}, \IF^{\textrm{trial}}) : \elasticmodule: \mathrm{d}{\strain} \leq 0$} 
                \State $\mathrm{d}\IVG = (\mathrm{d}\pstrain,\mathrm{d}\IV) = (\matrice{0},\vecteur{0})$, $\stress^{\textrm{new}} = \stress^{\textrm{trial}}$, $\epmodulenew = \elasticmodule$
            \Else
               \State  Solve the following constrained nonlinear system in $( \stress^{\textrm{new}},\IF^{\textrm{new}},\mathrm{d}\pstrain,\mathrm{d}\IV)$:
                \State  $\Phi(\stress^{\textrm{new}}, \IF^{\textrm{new}}) = 0$, \quad  $\partial_{\stress}\Phi(\stress^{\textrm{new}}, \IF^{\textrm{new}}) : \elasticmodule: ( \mathrm{d}{\strain} - \mathrm{d}{\pstrain}  )\leq 0$ \label{line_sept}
                \State  $\stress^{\textrm{new}} =  \elasticmodule : (\estrain + \mathrm{d}\strain - \mathrm{d}\pstrain), \quad \IF^{\textrm{new}} = \partial_{\IV}\Psi^p(\IV+  \mathrm{d}\IV)$ \label{line:sigma}
                \State $\mathrm{d}{\pstrain} = \Lambda(\stress^{\textrm{new}}, \IF^{\textrm{new}}) \, \partial_{\stress} \Phi(\stress^{\textrm{new}}, \IF^{\textrm{new}}), \quad \mathrm{d}{\IV} = - \Lambda(\stress^{\textrm{new}}, \IF^{\textrm{new}}) \, \partial_{\IF} \Phi(\stress^{\textrm{new}}, \IF^{\textrm{new}})$ \label{line_Lambda}
                \State and set $\epmodulenew = \elasticmodule - \dfrac{(\elasticmodule : \partial_{\stress} \Phi(\stress^{\textrm{new}}, \IF^{\textrm{new}})) \otimes (\elasticmodule : \partial_{\stress} \Phi(\stress^{\textrm{new}}, \IF^{\textrm{new}}))}{\partial_{\stress} \Phi(\stress^{\textrm{new}}, \IF^{\textrm{new}}):\elasticmodule:\partial_{\stress} \Phi(\stress^{\textrm{new}}, \IF^{\textrm{new}}) + \overline{H}(\stress^{\textrm{new}}, \IF^{\textrm{new}})}$ \label{line_Cep}
            \EndIf
            \State $\IVG^{\textrm{new}} = (\strain^{p, \textrm{new}}, \IV^{\textrm{new}}) = \IVG + \mathrm{d}\IVG$
            \State \Return $(\IVG^{\textrm{new}} , \stress^{\textrm{new}}, \epmodulenew)$
            \EndProcedure
    \end{algorithmic}
\end{algorithm}
Line~\ref{line:sigma} of Algorithm~\ref{algo::plasticity} shows that the increment in the stress tensor is given by $\mathrm{d}\stress = \elasticmodule : (\mathrm{d}\strain - \mathrm{d}\pstrain)$, and one then introduces the so-called consistent elastoplastic tangent modulus $\epmodulenew$ such that $\mathrm{d}\stress = \epmodulenew : \mathrm{d}\strain$; $\epmodulenew$ is a fourth-order tensor having minor and major symmetries. If the evolution is elastic, then $\epmodulenew = \elasticmodule$, and the consistent elastoplastic tangent modulus has two eigenvalues: $2\mu$ with multiplicity five and $3\lambda + 2 \mu$ with multiplicity one. Instead, if the evolution is plastic, then $\epmodulenew \neq  \elasticmodule $, and the eigenvalues of $\epmodulenew$, which are positive for strain-hardening plasticity (and non-negative for perfect plasticity) depend on the Helmholtz free energy $\Psi$ and  the yield function $\Phi$ (see \cite{Szabo1998} for details). For a finite incremental strain, the consistent elastoplastic tangent modulus generally differs from the so-called continuous elastoplastic tangent modulus which is obtained by letting the incremental strain tend to zero \cite{Simo1985a}.
\subsection{Example: combined linear isotropic and kinematic hardening with a von Mises yield criterion }\label{ss:combined_model}
An illustration of the plasticity model defined above is the combined linear isotropic and kinematic hardening model. The internal variables are $\IV :=(\pstrain ,p)$, where $p \geq 0$ is the equivalent plastic strain. We assume that the plastic part of the free energy is fully decoupled so that
\begin{equation}
\Psi^p(\IV) =  \frac{K}{2}\pstrain : \pstrain + \frac{H}{2} p^2.
\end{equation}
where $H \geq 0$, resp. $K \geq 0$, is the isotropic, resp. kinematic, hardening modulus. The associated thermodynamic forces $\IF :=(\matrice{\beta}, r)$ are the back-stress tensor $\matrice{\beta}=  K \pstrain$ and the internal stress $r = Hp$. Note that $\matrice{\beta}$ is a deviatoric tensor since $ \trace(\matrice{\beta})=0$. The perfect plastic model is retrieved by taking $H=K=0$. Concerning the yield function, we consider a $J_2$-plasticity model with a von Mises criterion:
\begin{equation}
\Phi(\stress, \IF)  = \sqrt{\frac{3}{2}} \Vert \dev( \stress  - \matrice{\beta}) \Vert_{\matrice{\ell}^2} - \sigma_{y,0} - r,
\end{equation}
where $\sigma_{y,0}$ is the initial yield stress, $\dev(\matrice{\tau}) := \matrice{\tau} - \frac{1}{d} \trace(\matrice{\tau})\matrice{I}$ is the deviatoric operator, and the Frobenius norm is defined as $\Vert\matrice{\tau}\Vert_{\matrice{\ell}^2} = \sqrt{ \matrice{\tau}  : \matrice{\tau} }$, for all $\matrice{\tau} \in \Reel^{d \times d}$.
The above model describes with a reasonable accuracy the behaviour of metals (see \cite{Lemaitre1994}).
\section{The Hybrid High-Order method}
\label{sec:HHO} 
\subsection{Discrete setting}
We consider a mesh sequence  $(\Th)_{h>0}$, where for each $h>0$, the mesh $\Th$ is composed of nonempty disjoint open polyhedra with planar faces such that $\overline{\Omega}_0 = \bigcup_{T\in\Th} \overline{T}$. The mesh-size is $h = \max_{T\in\Th} h_T$, where $h_T$ stands for the diameter of the cell $T$. A closed subset $F$ of $\overline{\Omega}_0$ is called a mesh face if it is a subset with nonempty relative interior of some affine hyperplane $H_F$ and \textup{(i)} if either there exist  two distinct mesh cells $T_-, \: T_+ \in \Th$ such that $F = \partial T_- \cap \partial T_+ \cap H_F$ (and $F$ is called an interface) or \textup{(ii)} there exists one mesh cell $T\in\Th$ such that $F = \partial T \cap \Gamma \cap H_F$ (and $F$ is called a boundary face). The mesh faces are collected in the set $\Fh$ which is further partitioned into the subset $\Fhi$ which is the collection of the interfaces and the subset $\Fhb$ which is the collection of the boundary faces. We assume that the mesh is compatible with the partition of the boundary $\Gamma$ into $\Bd$ and $\Bn$, so that we can further split the set $\Fhb$ into the disjoint subsets $\Fhbd$ and $\Fhbn$ with obvious notation. For all $T \in \Th$, $\FT$ is the collection of the mesh faces that are subsets of $\partial T$ and $\nT$ is the unit outward normal to $T$. We assume that the mesh sequence $(\Th)_{h>0}$ is shape-regular  in the sense specified in \cite{DiPEr:2015}, i.e., there is a matching simplicial submesh of $\Th$ that belongs to a shape-regular family of simplicial meshes in the usual sense of Ciarlet \cite{Ciarlet1978} and such that each mesh cell $T \in \Th$ (resp., mesh face $F \in \Fh$) can be decomposed in a finite number of sub-cells (resp., sub-faces) which belong to only one mesh cell (resp., to only one mesh face or to the interior of a mesh cell) with uniformly comparable diameter.

\begin{figure}
    \centering
    \includegraphics[scale=0.4]{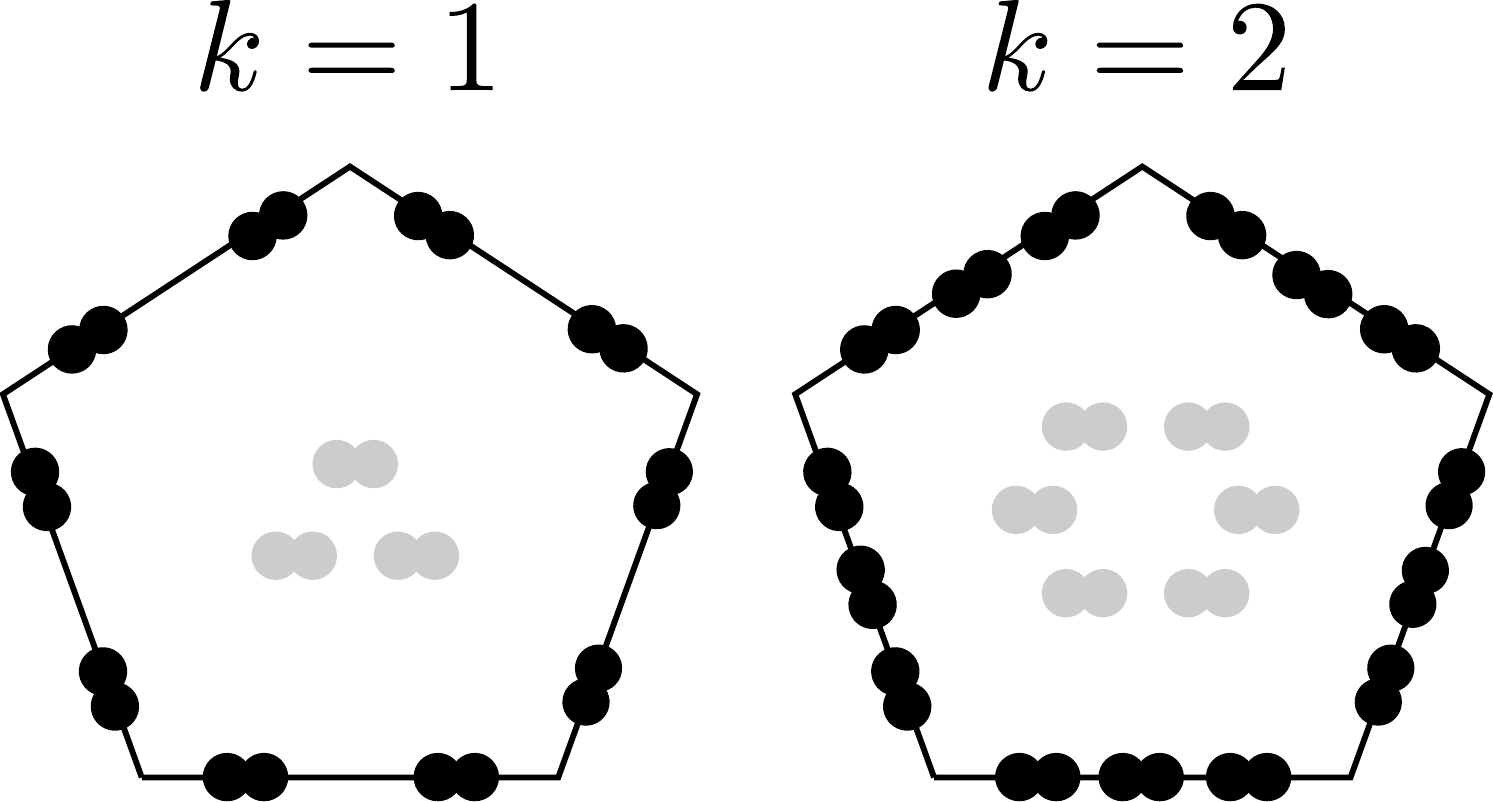} 
\caption{Face (black) and cell (gray) degrees of freedom in $\UkT$ for $k=1$ and $k=2$ in the two-dimensional case (each dot represents a degree of freedom which is not necessarily a point evaluation).}
\label{fig_HHO_dofs}
\end{figure}

Let $k \geq 1$  be a fixed polynomial degree. In each mesh cell $T\in\Th$, the local HHO unknowns consist of a pair $(\vT,\vdT)$, where the cell unknown  $\vT\in \Pkd(T; \Reel^d)$ is a vector-valued $d$-variate polynomial of degree at most $k$ in the mesh cell $T$, and $\vdT \in \PkF(\FT; \Reel^d) = \bigtimes_{F \in\FT} \PkF(F; \Reel^d)$ is a piecewise, vector-valued $(d-1)$-variate polynomial of degree at most $k$ on each face $F\in\FT$. We write more concisely that
\begin{equation}\label{local_dof}
(\vT,\vdT) \in \UkT := \Pkd(T;\Reel^d) \times \PkF(\FT;\Reel^d).
\end{equation}
The degrees of freedom are illustrated in Fig.~\ref{fig_HHO_dofs}, where a dot indicates one degree of freedom (which is not necessarily computed as a point evaluation). More generally, the polynomial degree $k$ of the face unknowns being fixed, HHO methods can be devised using cell unknowns that are polynomials of degree $l \in \{ k-1,k,k+1 \} \cap \mathbb{N}^{\star}$, (see \cite{CoDPE:2016}); these variants are briefly investigated numerically in \ifCM
\ref{ss:variant}.
\else
Appendix~\ref{ss:variant}.
\fi  We equip the space $\UkT$ with the following local discrete strain semi-norm:
\begin{equation} \label{eq:snorme}
\snorme[1,T]{( \vT, \vdT)}^2 := \normem[T]{\grads \vT}^2 + \normev[\dT]{\gamma_{\dT}^{\frac12}(\vT-\vdT)}^2,
\end{equation}
with the piecewise constant function $\gamma_{\dT}$ such that $\gamma_{\dT|F}=h_F^{-1}$ for all $F\in\FT$, where $h_F$ is the diameter of $F$. We notice that  $\snorme[1,T]{( \vT, \vdT)}=0$ implies that $\vT$ is a rigid-body motion and that $\vdT$ is the trace of $\vT$ on $\partial T$.
\subsection{Local symmetric strain reconstruction and stabilization}
\label{sec:grad_rec}
The first key ingredient in the devising of the HHO method is a local symmetric strain reconstruction in each mesh cell $T \in \Th$. This reconstruction is materialized by an operator $\EkT : \UkT \rightarrow \Pkd(T;\Msym)$ mapping onto the space composed of symmetric $\Reel^{d\times d}$-valued polynomials in $T$. The main reason for reconstructing the symmetric strain tensor in a larger space than the space $\grads \Pkpd(T; \Reel^d)$ originally introduced in \cite{DiPEr:2015} for the linear elasticity problem is that the reconstructed symmetric gradient of a test function acts against a discrete Cauchy stress tensor which is not in symmetric gradient form, see \cite[Section~4]{DiPietro2018} for further insight. For all $(\vT,\vdT) \in \UkT$, the reconstructed symmetric strain tensor $\EkT(\vT,\vdT) \in \Pkd(T;\Msym)$ is obtained by solving the following local problem: For all $\matrice{\tau} \in \Pkd(T;\Msym)$,
\begin{equation}\label{eq_reconstruction_grad}
\psm[T]{\EkT( \vT,\vdT)}{\matrice{\tau}} = \psm[T]{\grads{\vT}}{\matrice{\tau}} + \psv[\dT]{\vdT - \vT{}_{|\dT}}{\matrice{\tau} \, \nT}.
\end{equation}
Solving this problem entails choosing a basis of the polynomial space $\Pkd(T;\Reel)$ and inverting the associated mass matrix for each component of the tensor $\EkT(\vT,\vdT)$. 
The second key ingredient in the HHO method is a local stabilization operator that enforces weakly the matching between the faces unknowns and the trace of the cell unknowns. Following \cite{DiPEL:2014,DiPEr:2015}, the stabilization operator $\SdTk : \PkF(\FT;\Reel^d) \rightarrow \PkF(\FT;\Reel^d)$ acts on the difference $\vecteur{\theta} = \vdT - \vT{}_{|\dT} \in \PkF(\FT;\Reel^d)$ and is such that, for all  $\vecteur{\theta}\in \PkF(\FT;\Reel^d)$,
\begin{equation} \label{eq:stab}
\SdTk(\vecteur{\theta}) = \PikdT\big(\vecteur{\theta}-(\matrice{I}-\PikT)\DTkp(\vecteur{0},\vecteur{\theta})_{|\dT}\big),
\end{equation}
where $\PikT$ and $\PikdT$ denote, respectively, the $L^2$-orthogonal projectors onto $\Pkd(T;\Reel^d)$ and $\PkF(\FT;\Reel^d)$. Note that the right-hand side of~\eqref{eq:stab} can be rewritten as $\PikdT(\vdT - \vT{}_{|\dT}-(\matrice{I}-\PikT)\DTkp(\vT,\vdT)_{|\dT})$.
The local displacement reconstruction operator $\DTkp : \UkT \rightarrow \Poly_d^{k+1} (T; \Reel^d)$ is such that, for all $(\vT,\vdT) \in \UkT$, $\DTkp(\vT,\vdT) \in \Poly_d^{k+1}(T;\Reel^d)$ is obtained by solving the following local Neumann problem: For all $\vecteur{w} \in \Poly_d^{k+1} (T;\Reel^d)$,
\begin{equation}\label{eq_reconstruction_depl}
\psm[T]{\grads{\DTkp(\vT,\vdT)}}{\grads{\vecteur{w}}} = \psm[T]{\grads{\vT}}{\grads{\vecteur{w}}} + \psv[\dT]{\vdT - \vT{}_{|\dT}}{\grads{\vecteur{w}} \, \nT},
\end{equation}
together with the mean-value conditions $\int_T \DTkp(\vT,\vdT) \:dT= \int_T \vT \:dT$ and $\int_T \matrice{\nabla}^{ss} \DTkp(\vT,\vdT) \:dT= \int_{\partial T} \frac{1}{2}(\nT \otimes \vdT - \vdT \otimes  \nT) \:d\partial T$, where $\matrice{\nabla}^{ss}$ is the skew-symmetric part of the gradient operator. Comparing with~\eqref{eq_reconstruction_grad}, one readily sees that $\grads{\DTkp(\vT,\vdT)}$ is the $L^2$-orthogonal projection of $\EkT(\vT,\vdT)$ onto the subspace $\grads \Poly^{k+1}_d(T;\Reel^d)$. 
%
%
Following \cite[Lemma 4]{DiPEr:2015}, it is straightforward to establish the following stability and boundedness properties (the proof is omitted for brevity).
\begin{lemma}[Boundedness and stability]\label{lemma_stability_stab}
Let the symmetric strain reconstruction operator be defined by~\eqref{eq_reconstruction_grad} and the stabilization operator be defined by~\eqref{eq:stab}. Let $\gamma_{\dT}$ be defined below~\eqref{eq:snorme}. Then, we have the following properties: \textit{(i)} Boundedness: there exists $\alpha_\sharp>0$, uniform w.r.t.~$h$, so that, for all $T \in \Th$ and all  $(\vT,\vdT) \in \UkT$,
\begin{equation}
\bigg(\normem[T]{\EkT (\vT, \vdT)}^2 + \normev[\dT]{\gamma_{\dT}^{\frac12}\SdTk(\vdT - \vT{}_{|\dT})}^2\bigg)^{\frac12} \leq \alpha_\sharp \snorme[{1,T}]{(\vT, \vdT)}.
\end{equation}
\textit{(ii)} Stability: there exists $\alpha_\flat > 0$, uniform w.r.t.~$h$, so that, for all $T \in \Th$ and all $(\vT,\vdT) \in \UkT$,
\begin{equation}
\alpha_\flat \snorme[{1,T}]{(\vT, \vdT)} \leq \bigg(\normem[T]{\EkT (\vT, \vdT)}^2 + \normev[\dT]{\gamma_{\dT}^{\frac12}\SdTk(\vdT - \vT{}_{|\dT})}^2\bigg)^{\frac12}.
\end{equation}
\end{lemma}
As shown in \cite{DiPEr:2015}, the following important commuting property holds true:
\begin{equation}\label{eq:commuting}
\EkT(\vecteur{I}_{T,\dT}(\vecteur{v})) = \matrice{\Pi}^k_T(\grads\vecteur{v}),\qquad \forall \vecteur{v}\in H^1(T;\Reel^d),
\end{equation}
where the reduction operator $\vecteur{I}_{T,\dT} : H^1(T;\Reel^d)\rightarrow \UkT$ is defined so that $\vecteur{I}_{T,\dT}(\vecteur{v})=(\vecteur{\Pi}^k_T(\vecteur{v}),\vecteur{\Pi}^k_{\dT}(\vecteur{v}{}_{|\dT}))$. Proceeding as in \cite[Thm.~8]{DiPEr:2015}, one can show that for the linear elasticity problem and smooth solutions, the energy error  converges as $h^{k+1}|\vecteur{u}|_{\vecteur{H}^{k+2}(\Omega_0)}$. Finally, taking the trace in~\eqref{eq:commuting}, we infer that 
\begin{equation} \label{eq:trace_stab}
\trace\big(\EkT(\vecteur{I}_{T,\dT}(\vecteur{v}))\big) = \Pi_T^k(\divergence{\vecteur{v}}), \qquad \forall \vecteur{v}\in H^1(T;\Reel^d),
\end{equation}
which is the key commuting property used in \cite{DiPEr:2015} to prove robustness for quasi-incompressible linear elasticity. This absence of volumetric locking is confirmed in the numerical experiments performed in Section~\ref{sec::sec_numexp} in the nonlinear setting of incremental associative plasticity.
\subsection{Global discrete problem}
Let us now devise the global discrete problem. We set $\Pkd(\Th;\Reel^d) :=\bigtimes_{T \in\Th} \Pkd(T; \Reel^d)$ and $\PkF(\Fh;\Reel^d) := \bigtimes_{F \in\Fh} \PkF(F; \Reel^d)$. The global space of discrete HHO unknowns is defined as
\begin{equation} \label{eq:def_Ukh}
\Ukh := \Pkd(\Th;\Reel^d) \times \PkF(\Fh;\Reel^d).
\end{equation}
For an element $\vh \in \Ukh$, we use the generic notation $\vh = (\vTh,\vFh)$. For any mesh cell $T \in \Th$, we denote by $(\vT,\vdT)\in \UkT$ the local components of $\vh$ attached to the mesh cell $T$ and to the faces composing its boundary $\dT$, and for any mesh face $F\in\Fh$, we denote by $\vF$ the component of $\vh$ attached to the face $F$. The Dirichlet boundary condition on the displacement field can be enforced explicitly on the discrete unknowns attached to the boundary faces in $\Fhbd$. Letting $\PikF$ denote the $L^2$-orthogonal projector onto $\PkF(F;\Reel^d)$, we set
\begin{subequations}
\begin{align}
\Uknhd &:= \left\lbrace (\vTh, \vFh ) \in \Ukh \: \vert \: \vF = \PikF(\vu_{\textrm{D}}(t^n)), \; \forall F \in \Fhbd   \right \rbrace, \\
\Ukhz &:= \left\lbrace (\vTh, \vFh ) \in \Ukh \: \vert \: \vF = \vecteur{0}, \; \forall F \in \Fhbd   \right \rbrace.
\end{align}
\end{subequations}
Note that the map $\vh\mapsto (\sum_{T\in\Th} \snorme[1,T]{( \vT, \vdT)}^2)^{\frac12}$ defines a norm on $\Ukhz$ (see \cite[Prop.~5]{DiPEr:2015}).

A key feature of the present HHO method is that the discrete generalized internal variables are computed only at some quadrature points in each mesh cell. We introduce for all $T \in \Th$, the quadrature points $\Qp_T= (\Qp_{T,j})_{1 \leq j \leq m_Q}$, with $\Qp_{T,j} \in T$ for all $1 \leq j \leq m_Q$, and the quadrature weights $\vecteur{\Wp}_T= (\Wp_{T,j})_{1 \leq j \leq m_Q}$, with $\Wp_{T,j} \in \Reel$ for all $1 \leq j \leq m_Q$. We denote by $k_Q$ the order of the quadrature. Then, the discrete internal variables are sought in the space 
\begin{equation}
\IVGTh := \bigtimes_{T \in\Th}  \vecteur{X}^{m_Q},
\end{equation} 
that is, for all $T\in\Th$, the internal variables attached to $T$ form a vector $\IVG_T=(\IVG_T(\Qp_{T,j}))_{1\le j\le m_Q}$ with $\IVG_T(\Qp_{T,j})\in \vecteur{X}$ for all $1\le j\le m_Q$.

We can now formulate the global discrete problem. We will use the following notation for two tensor-valued functions defined on $T$:
\begin{equation}\label{eq_prods}
\psmd[T]{\matrice{s}}{\matrice{e}} := \sum_{j=1}^{m_Q} \Wp_{T,j} \, \matrice{s}(\Qp_{T,j}) : \matrice{e}(\Qp_{T,j}).
\end{equation}
We will also need to consider the case where we know the tensor $\matrice{\tilde s}$ only at the Gauss nodes (we use a tilde to indicate this situation), i.e., we have $\matrice{\tilde s} = (\matrice{\tilde s}(\Qp_{T,j}))_{1\le j\le m_Q} \in (\Reel^{d\times d})^{m_Q}$. In this case, we slightly abuse the notation by denoting again by $\psmd[T]{\matrice{\tilde s}}{\matrice{e}}$ the quantity equal to the right-hand side of~\eqref{eq_prods} with $\matrice{\tilde s}$ replacing $\matrice{s}$.
The global discrete problem consists in finding for any pseudo-time step $ 1 \leq n \leq N$, the pair of discrete displacements $(\uThn, \uFhn) \in \Uknhd$ and the discrete internal variables $\IVGh^n \in \IVGTh$ such that, for all $(\dvTh, \dvFh) \in \Ukhz$,
\begin{align}\label{discrete_problem_ptv}
&\sum_{T\in\Th} \psmd[T]{\tilde{\stress}^{n}}{\EkT(\dvT,\dvdT)}
+  \sum_{T\in\Th} \beta \psv[\dT]{\gamma_{\dT}\SdTk(\udTn- \uTn{}_{|\dT})}{\SdTk(\dvdT - \dvT{}_{|\dT})} \nonumber \\
&= \sum_{T\in\Th}  \psv[T]{ \loadext^n}{\dvT} + \sum_{F\in\Fhbn} \psv[F]{\Tn^n}{\dvF},
\end{align}
where for all $T \in \Th$ and all $1 \leq j \leq m_Q$,
\begin{multline}\label{discrete_problem_plast}
( \IVGT^n(\Qp_{T,j}), \tilde{\stress}^n(\Qp_{T,j}), \depmodulen(\Qp_{T,j}))  = \\
\textrm{PLASTICITY}( \IVGT^{n-1}(\Qp_{T,j}), \EkT(\uT^{n-1},\udT^{n-1})(\Qp_{T,j}),\EkT(\uT^{n},\udT^{n})(\Qp_{T,j})-\EkT(\uT^{n-1},\udT^{n-1})(\Qp_{T,j}) ),
\end{multline}%
with $(\uTh^{n-1}, \uFh^{n-1}) \in \Ukpnhd$ and $\IVGh^{n-1} \in \IVGTh$ given
either from the previous pseudo-time step or the initial condition.
Moreover, in the second line of~\eqref{discrete_problem_ptv}, the
stabilization employs
a weight of the form $\beta=2\mu \beta_0$ with $\beta_0>0$. In the original HHO method for linear elasticity \cite{DiPEr:2015}, the choice $\beta_0=1$ is considered. In the present setting, the choice for $\beta_0$ is further discussed in Section~\ref{sec:newton} and in \ifCM
\ref{ss:variant}.
\else
Appendix~\ref{ss:variant}.
\fi
\begin{remark}[Unstabilized HHO method]
An unstabilized HHO (uHHO) method has been considered in \citep{AbErPi2018} on affine simplicial meshes without hanging nodes for hyperelastic materials in finite deformations inspired by the stable dG methods without penalty parameters devised in \cite{John2016}; a more comprehensive treatment of unstabilized gradient reconstructions including polyhedral meshes can be found in \cite{DiPietro2018}. In the uHHO method from \citep{AbErPi2018}, the symmetric gradient is reconstructed in a larger space than $\Pkd(T; \Msym)$, typically $\Pkpd(T; \Msym)$, to achieve stability as in Lemma~\ref{lemma_stability_stab} with $\SdTk\equiv \vecteur{0}$. The price to be paid is a convergence rate of order $k$ for smooth solutions. In the present setting of elastoplasticity with small deformations, our numerical tests (not shown for brevity) indicate for $k=1$ less accurate  results for uHHO than for HHO with stabilization, and for $k=2$, the results are of comparable accuracy. The CPU costs are more or less comparable since the time saved by avoiding the stabilization for uHHO is compensated  by the need to reconstruct the strain in a larger space.
\end{remark}
\subsection{Discrete principle of virtual work}
The discrete problem~\eqref{discrete_problem_ptv} expresses the principle of virtual work at the global level, and following the ideas introduced in \cite{CoDPE:2016} (see also \cite{BoDPS:2017,AbErPi2018}), it is possible to infer a local principle of virtual work in terms of face-based discrete tractions that comply with the law of action and reaction. 

Let $\SdTks : \PkF(\FT;\Reel^d) \rightarrow \PkF(\FT;\Reel^d)$ be the adjoint operator of $\SdTk$ with respect to the $\vecteur{L}^2(\dT)$-inner product so that we have $\psv[\dT]{\gamma_{\dT}\SdTk(\vecteur{\theta})}{\SdTk(\vecteur{\zeta})} = \psv[\dT]{\SdTks(\gamma_{\dT}\SdTk(\vecteur{\theta}))}{\vecteur{\zeta}}$ (recall that the weight $\gamma_{\dT}$ is piecewise constant on $\dT$). 
Let $\PikTmd: (\Reel^{d \times d})^{m_Q} \rightarrow \Pkd(T;\Reel^{d \times d} )$ denote the $L^2_Q$-orthogonal projector such that for all $\tilde{\matrice{s}} \in (\Reel^{d \times d})^{m_Q}$, $\psm[T]{\PikTmd(\tilde{\matrice{s}})}{\matrice{e}} = \psmd[T]{\tilde{\matrice{s}}}{\matrice{e}}$ for all $\matrice{e} \in \Pkd(T; \Reel^{d \times d})$. Finally, for any pseudo-time step $1\le n\le N$ and all $T\in\Th$, let us define the discrete traction:
\begin{equation}\label{eq:traction_stab}
\vecteur{T}_{T}^{n} := \PikTmd(\dstress_T^{n})\SCAL\nT
+ \beta \SdTks(\gamma_{\dT}\SdTk(\udTn- \uTn{}_{|\dT}))\in \PkF(\FT; \Reel^d),
\end{equation} 
where $\dstress_T^n =  (\dstress_{T}^n(\Qp_{T,j}))_{1 \leq j \leq m_Q} \in( \Msym)^{m_Q}$ with $\dstress_{T}^n(\Qp_{T,j}) = \elasticmodule : (\EkT(\uTn,\udTn)(\Qp_{T,j}) - \tilde{\strain}^{p,n}_T(\Qp_{T,j}))$ for all $1 \leq j \leq m_Q$.
\begin{lemma}[Equilibrated tractions] \label{lem:equil_sHHO}
Assume that $k_Q\ge 2k$. Then, for any pseudo-time step $1 \leq n \leq N$, the following local principle of virtual work holds true for all $T\in\Th$:
\begin{equation} \label{eq:pwk}
\psmd[T]{\dstress^n_T}{\grads{\dvT}}-
\psv[\dT]{\vecteur{T}_{T}^n}{\dvT} = \psv[T]{ \loadext^n}{\dvT},
\qquad \forall \dvT \in \Pkd(T;\Reel^d),
\end{equation}
where the discrete tractions $\vecteur{T}_{T}^n$ defined by~\eqref{eq:traction_stab} satisfy the following law of action and reaction for all $F\in\Fhi\cup\Fhbn$:
\begin{subequations}\label{eq:balance}
\begin{alignat}{2}
&\vecteur{T}_{T_-|F}^n + \vecteur{T}_{T_+|F}^n = \vecteur{0},
&\quad&\text{if $F\in\Fhi$ with $F=\partial T_- \cap \partial T_+ \cap H_F$},\\
&\vecteur{T}_{T|F}^n  = \PikF(\Tn^n),&\quad&\text{if $F\in\Fhbn$ with $F=\partial T\cap \Bn\cap H_F$}.
\end{alignat}
\end{subequations}  
\end{lemma}
The proof is postponed to 
\ifCM
\ref{proof_dw}.
\else
Appendix~\ref{proof_dw}.
\fi
\subsection{Nonlinear solver}
\label{sec:newton}
The nonlinear problem \eqref{discrete_problem_ptv}-\eqref{discrete_problem_plast} arising at any pseudo-time step $1 \leq n \leq N$ is solved using a Newton's method. Given  $(\uTh^{n-1}, \uFh^{n-1}) \in \Ukpnhd$ and $\IVGh^{n-1} \in \IVGTh$ from the previous pseudo-time step or the initial condition, the Newton's method is initialized by setting $(\uTh^{n,0}, \uFh^{n,0})=(\uTh^{n-1}, \uFh^{n-1})$ up to the update of the Dirichlet condition and $\IVGh^{n,0}= \IVGh^{n-1}$. Then, at each Newton's step $i\ge0$, one computes the incremental displacement $(\duTh^{n,i}, \duFh^{n,i}) \in \Ukhz$ and updates the discrete displacement as $(\uTh^{n,i+1} \uFh^{n,i+1}) = (\uTh^{n,i}, \uFh^{n,i})+(\duTh^{n,i}, \duFh^{n,i})$. The linear system of equations to be solved is
\begin{align}\label{eq_stiffness_matrix}
& \hphantom{+} \sum_{T\in\Th} \psmd[T]{\depmoduleni: \EkT(\duT^{n,i},\dudT^{n,i})}{\EkT(\dvT,\dvdT)} \nonumber \\ 
&+  \sum_{T\in\Th} \beta \psv[\dT]{\gamma_{\dT}\SdTk(\dudT^{n,i} - 
\duT^{n,i}{}_{|\dT})}{\SdTk(\dvdT - \dvT{}_{|\dT})} \nonumber \\ 
&= -R_h^{n,i}(\dvTh,\dvFh),
\end{align}
for all $(\dvTh,\dvFh)\in\Ukhz$, where for all $T \in \Th$ and all $1 \leq j \leq m_Q$,
\begin{align}
 &(\IVGT^{n,i}(\Qp_{T,j}), \dstress^{n,i}(\Qp_{T,j}), \depmoduleni(\Qp_{T,j})) 
= \textrm{PLASTICITY}(\IVG_{T,j}^{n-1},\matrice{e}_{T,j}^{n-1},\matrice{e}_{T,j}^{n,i}-\matrice{e}_{T,j}^{n-1}),
\end{align}
with $\IVG_{T,j}^{n-1} = \IVGT^{n-1}(\Qp_{T,j})$, $\matrice{e}_{T,j}^{n,i} = \EkT(\uT^{n,i},\udT^{n,i})(\Qp_{T,j})$, 
$\matrice{e}_{T,j}^{n-1} = \EkT(\uT^{n-1},\udT^{n-1})(\Qp_{T,j})$, and the residual term
\begin{align}
R_h^{n,i}(\dvTh, \dvFh) 
={}& \sum_{T\in\Th} \psmd[T]{\dstress^{n,i}}{\EkT(\dvT,\dvdT)} - \sum_{T\in\Th}  \psv[T]{ \loadext^n}{\dvT} - \sum_{F\in\Fhbn} \psv[F]{\Tn^n}{\dvF}  \nonumber \\
&+  \sum_{T\in\Th} \beta \psv[\dT]{\gamma_{\dT}\SdTk(\udT^{n,i} - \uT^{n,i}{}_{|\dT})}{\SdTk(\dvdT - \dvT{}_{|\dT})}.
\end{align}
The assembling of the stiffness matrix resulting from the left-hand side of~\eqref{eq_stiffness_matrix} is local (and thus fully parallelizable). The discrete internal variables $\IVGh^{n} \in \IVGTh$ are updated at the end of each pseudo-time step.

For strain-hardening plasticity, the consistent elastoplastic tangent modulus $\epmodule$ is symmetric positive-definite. Let us set $\theta_{\Th, Q} := \min_{(T,j) \in \Th\times \{1,\ldots,m_Q\}} \theta^{\min}(\depmodule(\Qp_{T,j}))$, where $\theta^{\min}(\tenseur{4}{{\mathbb{M}}})$ denotes the smallest eigenvalue of the symmetric fourth-order tensor $\tenseur{4}{{\mathbb{M}}}$. The following result shows that the linear system~\eqref{eq_stiffness_matrix} arising at each Newton's step is coercive under the simple choice $\beta_0>0$ on the stabilization parameter for strain-hardening plasticity.

\begin{theorem}[Coercivity]\label{th::coer}
Assume that $k_Q\ge 2k$ and that all the quadrature weights are positive. Moreover, assume that  the plastic model is strain-hardening. Then, the linear system~\eqref{eq_stiffness_matrix} in each Newton's step is coercive for all $\beta_0 > 0$, i.e., there exists $C_{\rm ell}>0$, independent of $h$, such that for all $(\vTh, \vFh) \in \Ukhz$,
\begin{multline}\label{eq_stable}
\sum_{T\in\Th} \psmd[T]{\depmodule: \EkT(\vT,\vdT)}{\EkT(\vT,\vdT)}+  \sum_{T\in\Th} \beta \psv[\dT]{\gamma_{\dT}\SdTk(\vdT- \vT{}_{|\dT})}{\SdTk(\vdT - \vT{}_{|\dT})} \\
\geq C_{\rm ell} \min\left(\beta_0, \frac{\theta_{\Th, Q}}{2\mu} \right) 2\mu \sum_{T\in\Th} \snorme[{1,T}]{(\vT, \vdT)}^2.
\end{multline}
\end{theorem}
The proof is postponed to 
\ifCM
\ref{proof_coer}.
\else
Appendix~\ref{proof_coer}. 
\fi
\begin{remark}
Theorem~\ref{th::coer} remains valid if  the assumption on the strain-hardening plasticity model is replaced by the weaker assumption that $ \theta_{\Th, Q}>0$ which is verified if the consistent elastoplastic tangent modulus is symmetric positive-definite.
\end{remark}
A reasonable choice of the stabilization parameter appears to be $\beta_0 \geq \max(1,\frac{\ds \theta_{\Th, Q}}{2\mu})$ because $\beta_0 = 1$ is a natural choice for the linear elasticity problem (see \cite{DiPEr:2015}) and the choice $\beta_0 \geq \frac{ \ds\theta_{\Th, Q}}{2\mu}$ allows one to adjust the stabilization parameter if the evolution is plastic. We investigate numerically the choice of $\beta_0$ in \ifCM
\ref{ss:variant}.
\else
Appendix~\ref{ss:variant}.
\fi For the combined linear isotropic and kinematic plasticity model with a von Mises yield criterion, the suggested choice leads to $\beta_0 \geq 1$ since the smallest eigenvalue $\theta_{\Th, Q}$ is such that
$\frac{\ds \theta_{\Th, Q}}{2\mu} =  \frac{\ds \overline{H}}{\mu + \overline{H}} < 1$ for the continuous elastoplastic tangent modulus.
\subsection{Implementation and static condensation}
\label{sec:implement}
As is classical with HHO methods \cite{DiPEr:2015, DiPEL:2014}, and more generally with hybrid approximation methods, the cell unknowns $\duT^{n,i}$ in~\eqref{eq_stiffness_matrix} can be eliminated locally by using a static condensation (or Schur complement) technique. Indeed, testing~\eqref{eq_stiffness_matrix} against the function $((\dvT\delta_{T,T'})_{T'\in\Th},(\vecteur{0})_{F\in\Fh})$ with Kronecker delta $\delta_{T,T'}$ and $\dvT$ arbitrary in $\Pkd(T;\Reel^d)$, one can express, for all $T\in\Th$, the cell unknown $\duT^{n,i}$ in terms of the local face unknowns collected in $\dudT^{n,i}$. As a result, the static condensation technique allows one to reduce~\eqref{eq_stiffness_matrix} to a linear system in terms of the face unknowns only. The reduced system is of size $N_{\Fh} \times  d{k+d-1\choose d-1}$, where $N_{\Fh}$ denotes the number of mesh faces (Dirichlet boundary faces can be eliminated by enforcing the boundary condition explicitly). In the reduced system, each mesh face is connected to neighbouring faces that share a mesh cell with the face in question.  Note that static condensation can improve the condition number of the global stiffness matrix; we refer the reader to \cite[Section~4.3]{AbErPi2018} for numerical results in the case of HHO methods for hyperelastic materials. Since the behavior integration is performed at the cell level, the same procedure as for cG methods can be used to deal with a large variety of behavior laws. One salient example is the standard radial return mapping \cite{Simo1998} (see also \cite{Auricchio2003, Szabo2009}) that will be used in the numerical examples of Section~\ref{sec::sec_numexp} to solve the nonlinear problem in Algorithm~\ref{algo::plasticity}. 

The implementation of HHO methods is realized using the open-source library \texttt{DiSk++} \cite{CicDE:2017Submitted} which provides generic programming tools for the implementation of HHO methods and is available online\footnote{ \texttt{https://github.com/wareHHOuse/diskpp}}. The data structure requires access to faces and cells as in standard dG or HDG codes. The reconstruction and stabilization operators are built locally at the cell level using scaled translated monomials to define the basis functions (see \cite[Section 3.2.1]{CicDE:2017Submitted} for more details). If memory is not a limiting factor, it is computationally effective to compute these operators once and for all in all the mesh cells, and to re-use them at each Newton's step. The \texttt{DiSk++} library employs different quadratures depending on the type of the mesh cells. On segments, standard Gauss quadrature is used. On quadrilaterals and hexahedra, the quadrature is obtained by tensorizing the one-dimensional Gauss quadrature. On triangles and tetrahedra, the Dunavant \cite{Dunavant1985} and Grundmann--Moeller \cite{Grundmann1978} quadratures are used, respectively. Polyhedral cells are split into sub-simplices, and the integration is performed in each sub-simplex separately. The linear algebra operations are realized using the Eigen library and the global linear system (involving face unknowns only) is solved with PardisoLU from the MKL library (alternatively, iterative solvers are also applicable). Finally, the Dirichlet boundary conditions are enforced strongly on the face unknowns as described above. 
\section{Numerical examples}\label{sec::sec_numexp}
The goal of this section is to evaluate the proposed HHO method on two- and three-dimensional test cases from the literature: a sphere under internal pressure, a quasi-incompressible Cook's membrane, a perforated strip subjected to uniaxial extension, and a cube under compression. We compare the results produced by the HHO method to the analytical solution whenever available or to numerical results obtained using the industrial open-source FEM code \CA \cite{CodeAster}. In this case, we consider a quadratic cG formulation, referred to as T2 or Q2 depending on the mesh, and a three-field formulation in which the unknowns are the displacement, the pressure and the volumetric strain fields referred to as UPG \cite{AlAkhrass2014}; in the UPG method, the displacement field is quadratic, whereas both the pressure and the volumetric strain fields are linear. The T2 and Q2 methods are
known to present volumetric locking due to plastic incompressibility,
whereas the UPG method is known to be robust but costly. 
Numerical results obtained using the UPG method on a very fine grid are used as a reference solution whenever an analytical solution is not available. Moreover, we also investigate the behavior of 
the HHO method on general meshes.
The combined linear isotropic and kinematic plasticity model with a von Mises yield criterion described in Section~\ref{ss:combined_model} is used for all the test cases. Strain-hardening plasticity is considered for the two-dimensional cases, i.e., the quasi-incompressible Cook's membrane and the perforated strip under uniaxial traction, whereas  perfect plasticity  is considered for the three-dimensional cases, i.e., the sphere under internal pressure and the cube under compression. Moreover, for the two-dimensional test cases, we assume additionally a plane strain condition. In the numerical experiments reported in this section, the stabilization parameter is taken to be $\beta = 2 \mu$ ($\beta_0=1$), and all the quadratures use positive weights. We employ the notation HHO($k$) when using face (and cell) polynomials of order $k$. All the tests are run sequentially on a 3.4 Ghz Intel Xeon processor with 16 Gb of RAM.

\subsection{Sphere under internal pressure}\label{ss::sphere}
\begin{figure}[htbp]
    \centering
    \subfloat[]{
    \label{fig::sphere_mesh}
        \centering 
        \includegraphics[scale=0.35]{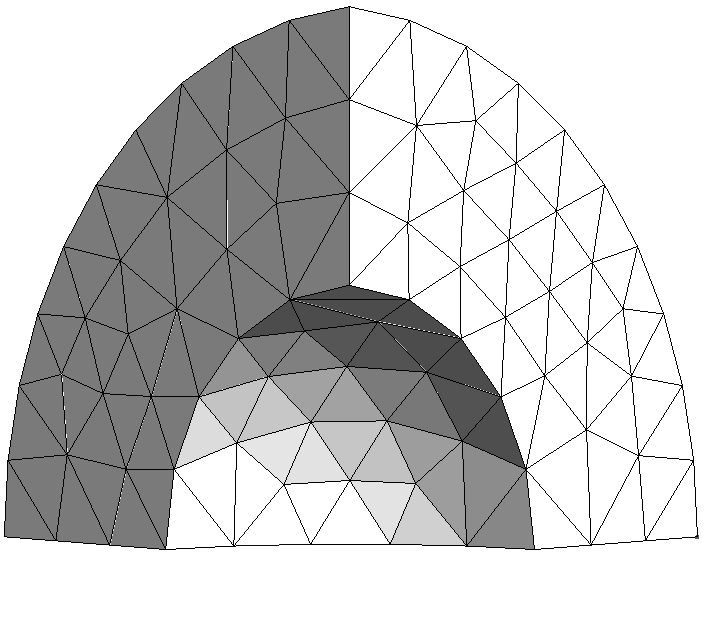} 
  }
    ~ 
    \subfloat[]{
        \centering
	    \includegraphics[scale=1.0]{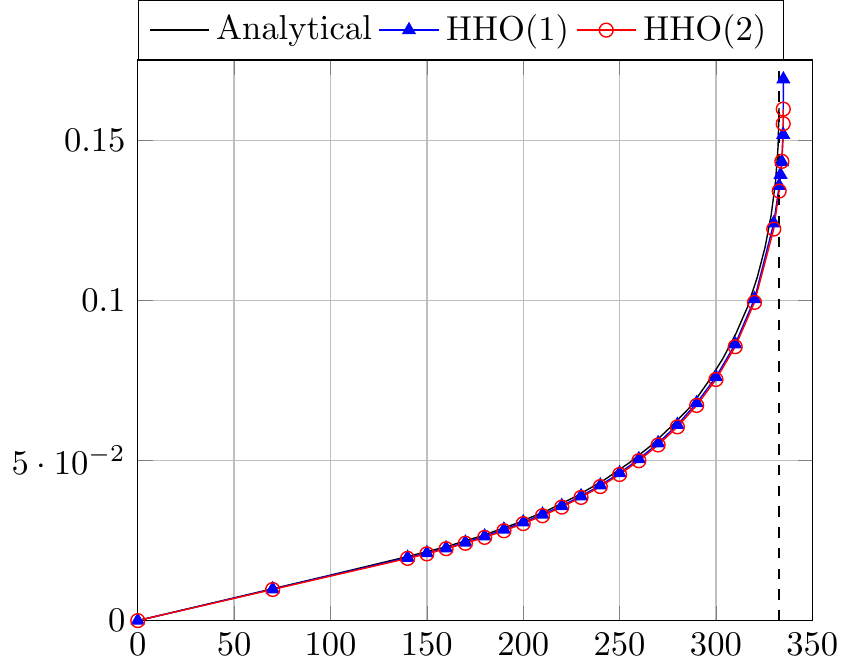}
	    \label{fig::sphere_load}
    }
    \caption{Sphere under internal pressure: (a) Mesh in the reference configuration composed of 506 tetrahedra. (b) Average radial displacement at the outer surface $(\mm)$ vs. applied pressure $(\MPa)$; the dashed line indicates the theoretical limit load.}
\end{figure}

\begin{figure}[htbp]
    \centering
    \subfloat[$\sigma_{rr}~(\MPa)$ vs. $r~(\mm)$]{
    \label{fig::sphere_srr}
        \centering 
        \includegraphics[scale=0.92]{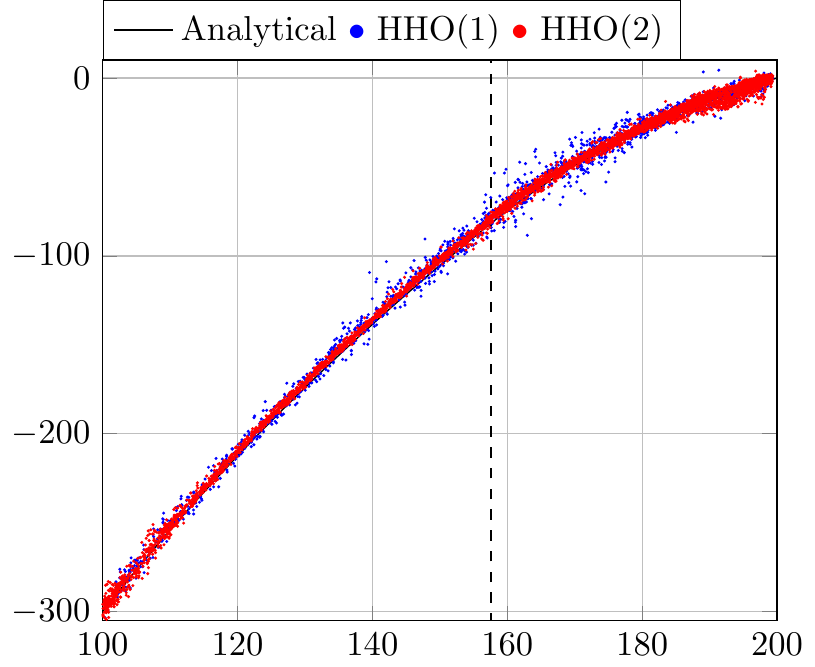} 
  }
    ~ 
    \subfloat[$\sigma_{\theta \theta}~(\MPa)$ vs. $r~(\mm)$]{
        \centering
	    \includegraphics[scale=0.92]{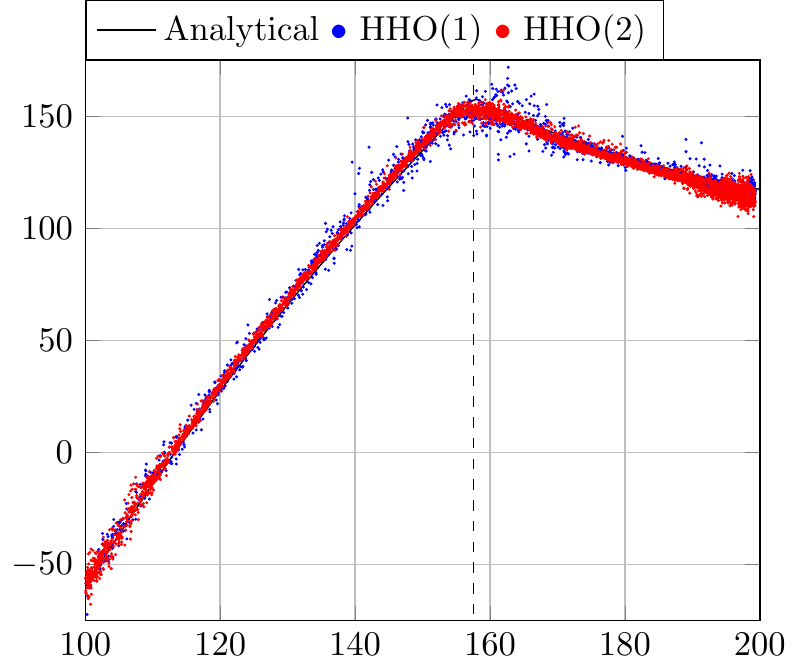}
	    \label{fig::sphere_soo}
    }
    \caption{Sphere under internal pressure: radial (left) and hoop (right) components of the stress tensor ($\MPa$)  vs. $r~(\mm)$ for HHO(1) and HHO(2) at all the quadrature points and for $P=300~\MPa$ (the dashed line corresponds to the transition between the plastic zone and the elastic zone).}
    \label{fig::sphere_stress}
\end{figure}

\begin{figure}[htbp]
    \centering
    \subfloat[]{
    \label{fig::sphere_trace_hho}
        \centering 
        \includegraphics[scale=0.9]{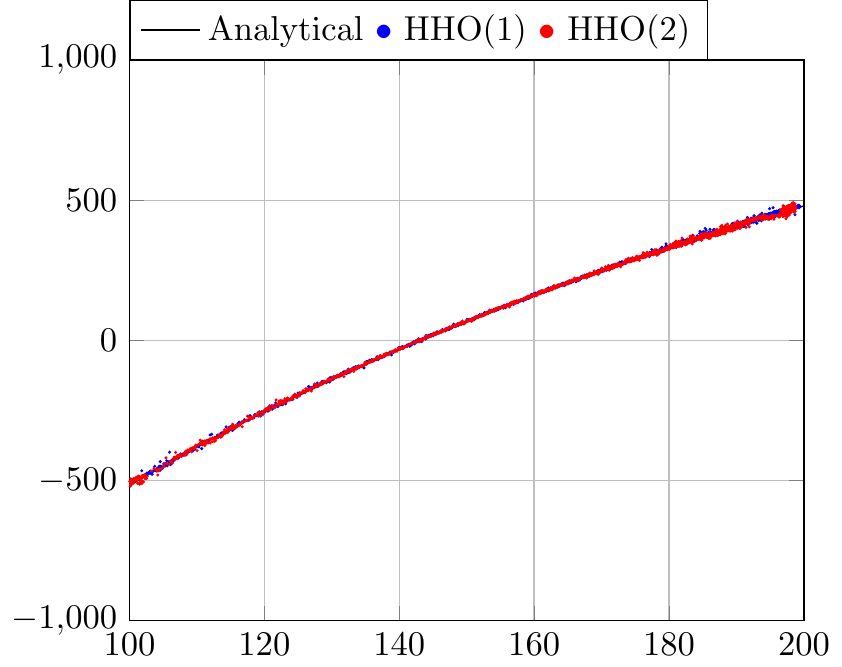} 
  }
    ~ 
    \subfloat[]{
        \centering
	    \includegraphics[scale=0.9]{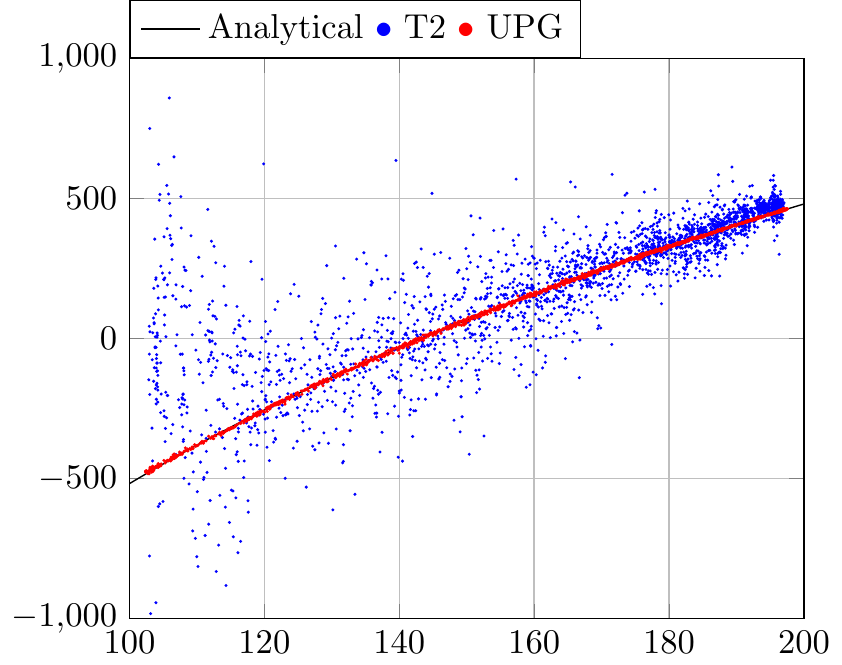}
	    \label{fig::sphere_trace_aster}
    }
    \caption{Sphere under internal pressure: trace of the stress tensor $(\MPa$) vs. $r~(\mm)$ at all the quadrature points and for the limit load; (a) HHO(1) and HHO(2), (b) T2 and UPG.}
    \label{fig::sphere_trace}
\end{figure}

This first benchmark consists of a sphere under internal pressure for which an analytical solution is known (see \cite[Section~7.5.2]{Neto2011}). The sphere has an inner radius $R_{in}=100~\mm$ and an outer radius $R_{out}=200~\mm$. An internal radial pressure $P$ is imposed. The material parameters adopted are those of  \cite{Neto2011}: a Young modulus $E =210~\GPa$, a Poisson ratio $\nu=0.3$, and an initial yield stress $\sigma_{y,0} = 240~\MPa$. For symmetry reasons, only one-eighth of the sphere is discretized, and the mesh is composed of 506 tetrahedra, see Fig.~\ref{fig::sphere_mesh}. The simulation is performed until the limit load corresponding to an internal pressure $P_{lim} \simeq 332.71~\MPa$ is reached. Numerically, this limit load is reached when the Newton solver stops converging. The load-deflection curves are plotted for HHO methods in Fig.~\ref{fig::sphere_load} showing that both HHO(1) and HHO(2) produce numerical results in very good agreement with the analytical solution. For this test case, we do not expect that HHO(2) will deliver much more accurate solutions than HHO(1) since the geometry is discretized using tetrahedra with planar faces.
The computed radial $(\sigma_{rr})$ and hoop $(\sigma_{\theta \theta})$ components of the stress tensor are shown in Fig.~\ref{fig::sphere_stress} at all the quadrature points for $P=300~\MPa$. For both HHO(1) and HHO(2), the computed stresses are very close to the analytical solution. The error on the stresses is slightly larger for HHO(1) than for HHO(2) near the transition between the elastic zone and the plastic zone (indicated by a dashed line at the radius $R_p = 157.56~\mm$). Finally, the trace of the stress tensor is compared for HHO, UPG and cG methods in Fig.~\ref{fig::sphere_trace} at all the quadrature points for the limit load. A sign of locking is the presence of strong oscillations of the trace of the stress tensor. Thus, we notice that the quadratic element T2 locks, whereas HHO and UPG do not present any sign of locking and produce results that are very close to the analytical solution.
\subsection{Quasi-incompressible Cook's membrane} \label{ss::cook}
\begin{figure}[htbp]
    \centering
    \subfloat[]{
    \label{fig::cook_geom}
        \centering 
        \includegraphics[scale=0.35]{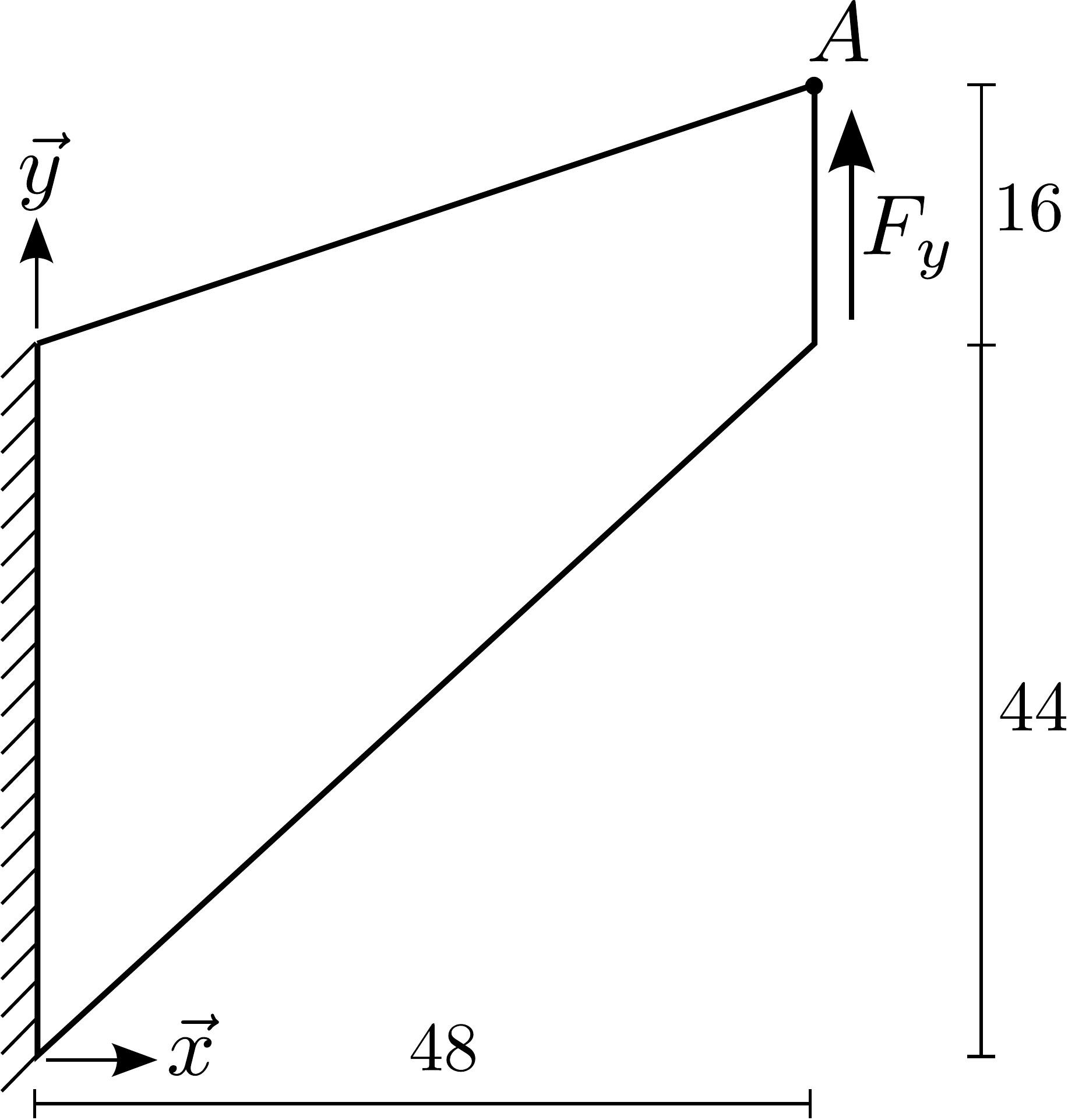} 
  }
    ~ 
    \subfloat[]{
        \centering
	    \includegraphics[scale=1.0]{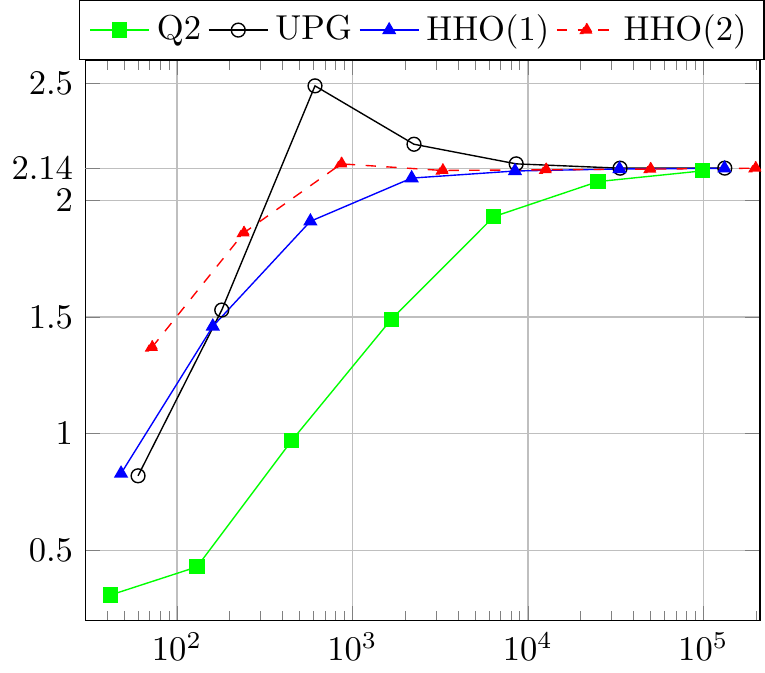}
	    \label{fig::cook_depl}
    }
    \caption{Quasi-incompressible Cook's membrane: (a) Geometry and boundary conditions (dimensions in $\mm$). (b) Convergence of the vertical displacement of the point $A$ (in $\mm$) vs. the number of degrees of freedom for Q2, UPG, HHO(1), and HHO(2).}
\end{figure}

\begin{figure}
\centering
\includegraphics[scale=0.28]{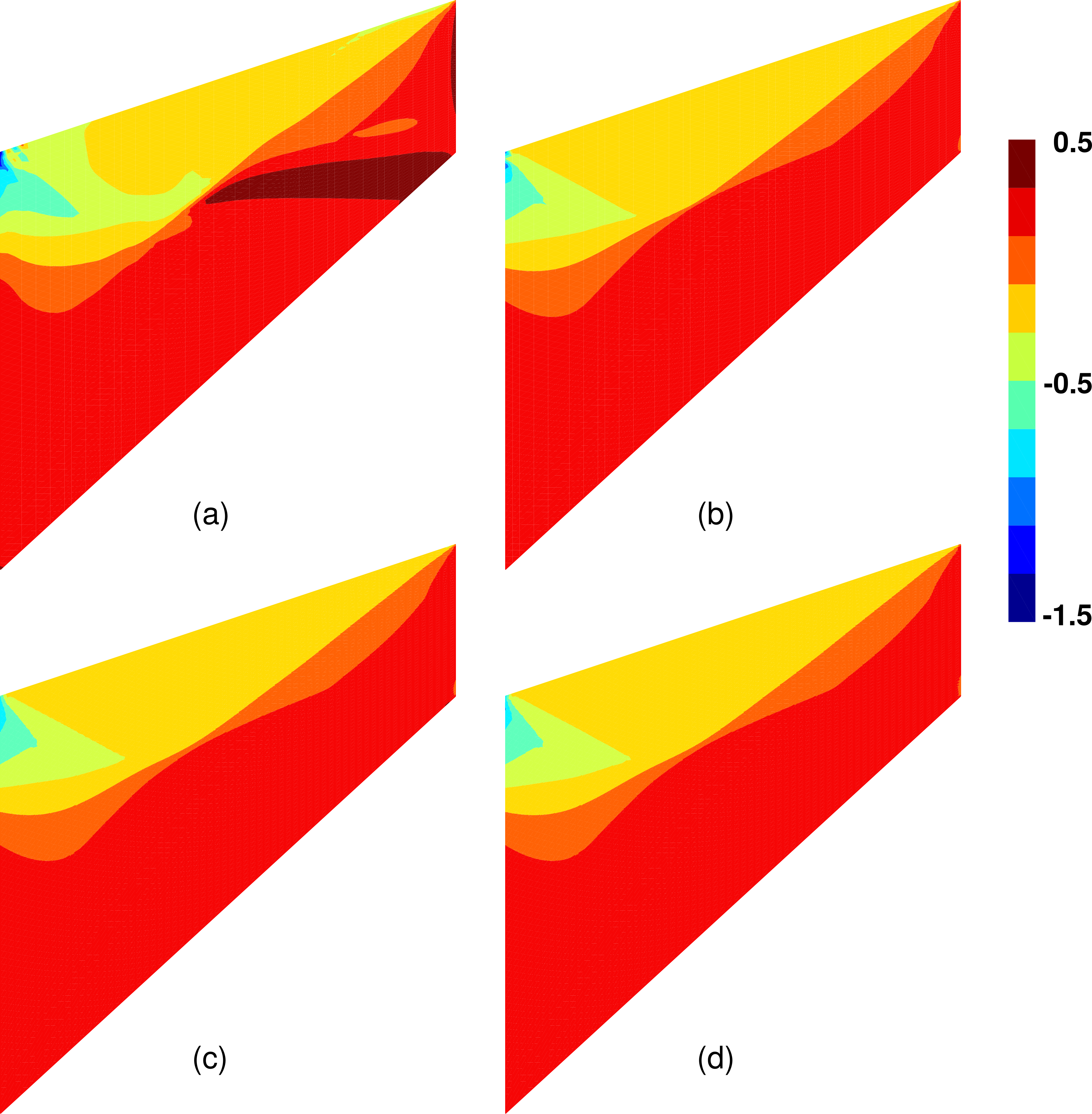}
\caption{Quasi-incompressible Cook's membrane: trace of the stress tensor $(\MPa)$ on the reference configuration for a $64 \times 64$ quadrangular mesh; (a) Q2 (b) UPG (c) HHO(1) (d) HHO(2).}
\label{fig::cook_trace}
\end{figure}

We consider the quasi-incompressible Cook's membrane problem which is a well known bending-dominated test case (see for example \cite[Section~6.2.3]{Simo1990} or \cite{Chiumenti2004}). It consists of a tapered panel, clamped on one side, and subjected to a vertical load $F_y = 1.8~\textrm{N}$ on the opposite side, as shown in Fig.~\ref{fig::cook_geom}. The material parameters are a Young modulus $E = 70~\MPa$, a Poisson ratio $\nu=0.4999$, an initial yield stress $\sigma_{y,0} = 0.243~\MPa$, an isotropic hardening modulus $H=0.135~\MPa$ and a kinematic hardening modulus $K=0~\MPa$. The simulation is performed in twenty uniform increments of the load and with a sequence of refined quadrangular meshes such that each side contains $2^N$ edges, $1\leq N \leq 7$. The vertical displacement of the point $A$ is plotted in Fig.~\ref{fig::cook_depl} for the HHO(1), HHO(2), cG, and UPG methods. As expected when comparing the number of degrees of freedom, the quadratic cG formulation Q2 has the slower convergence, HHO(1) converges slightly faster than UPG, and HHO(2) outperforms all the other methods. Moreover, we show in Fig.~\ref{fig::cook_trace} the trace of the stress tensor. The cG formulation Q2 presents strong oscillations that confirm the presence of volumetric locking, contrary to the HHO and UPG methods which deliver similar and smooth results.
\subsection{Perforated strip subjected to uniaxial extension}\label{ss::plaque}
\begin{figure}[htbp]
    \centering
    \subfloat[]{
    \label{fig:plaque_geom}
        \centering 
        \includegraphics[scale=0.7]{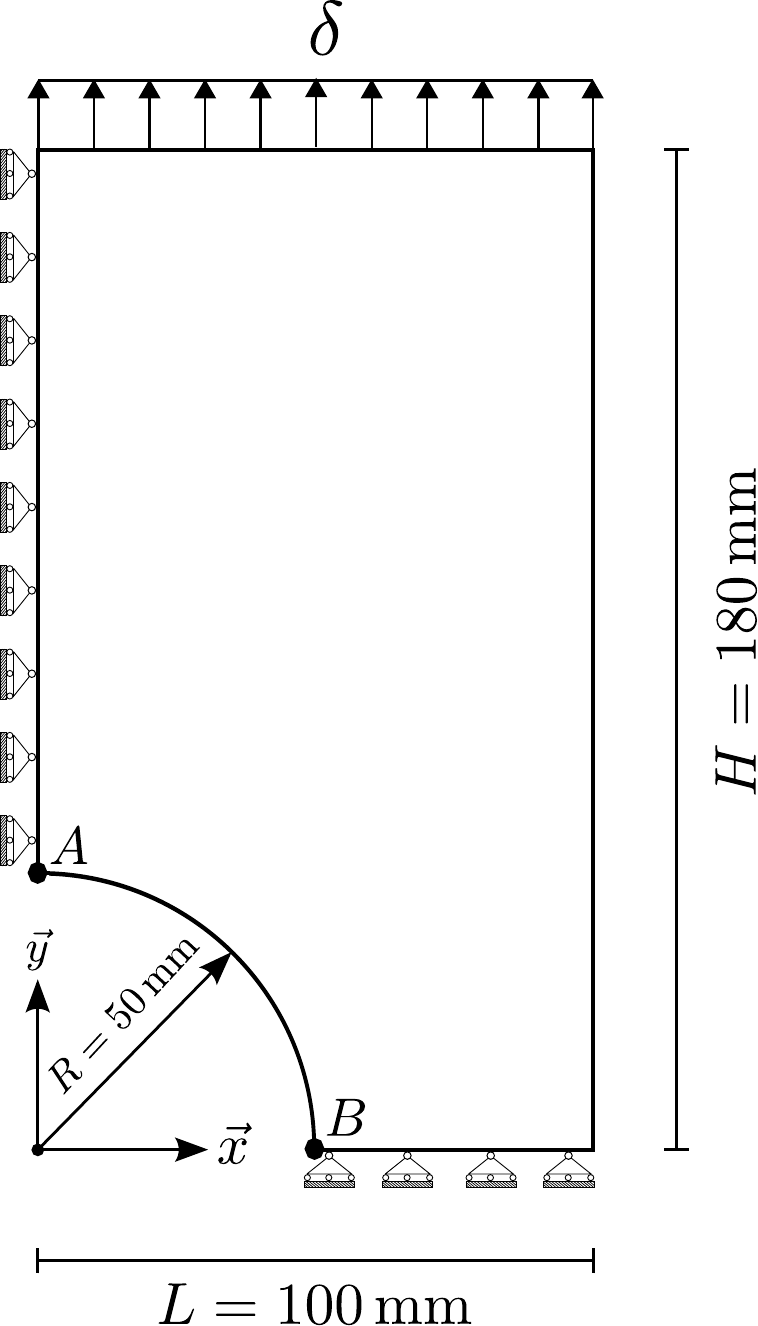} 
  }
    ~ 
    \subfloat[]{
        \centering
	    \raisebox{12mm}{\includegraphics[scale=0.34]{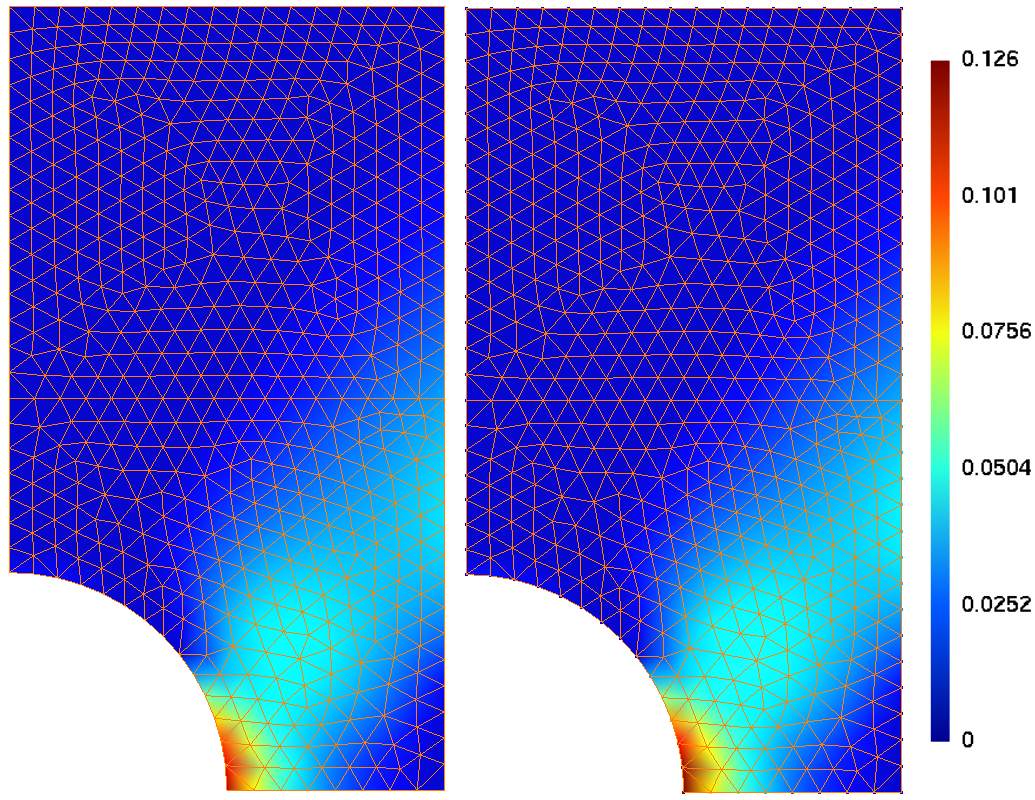}}
	    \label{fig::plaque_p}
    }
    \caption{Perforated strip: (a) Geometry. (b) Equivalent plastic strain $p$ for a triangular mesh with UPG (left) and HHO(2) (right);  there are 5,542 degrees of freedom  for UPG and 9,750 for HHO(2).}
\end{figure}

\begin{figure}[htbp]
    \centering
    \subfloat[Point $A$]{
        \centering 
        \includegraphics[scale=0.9]{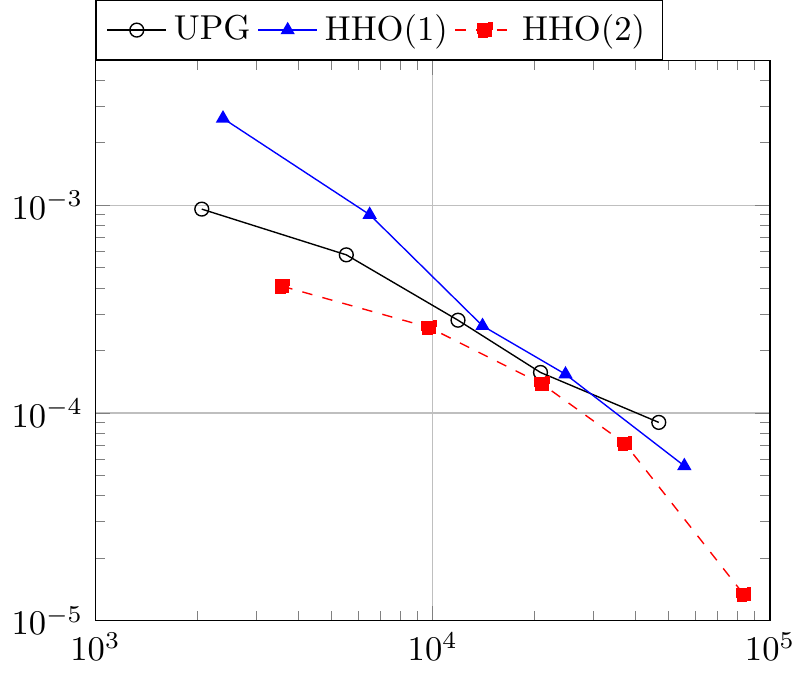} 
  }
    ~ 
    \subfloat[Point $B$]{
        \centering
	    \includegraphics[scale=0.9]{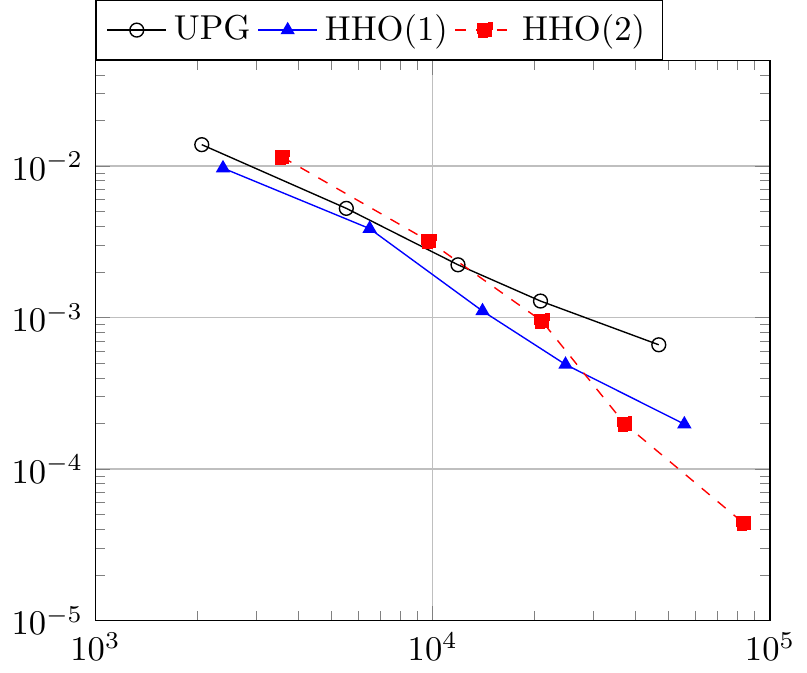}
    }
    \caption{Perforated strip: Relative displacement error at the points $A$ (left) and $B$ (right) vs. the number of degrees of freedom for UPG, HHO(1) and HHO(2).}
    \label{fig::plaque_error}
\end{figure}

We consider a strip of width $2L=200~\mm$ and height $2H=360~\mm$. The strip is perforated in its middle by a circular hole of radius $R=50~\mm$, and  is subjected to a uniaxial extension $\delta=5~\mm$ at its top and bottom ends. For symmetry reasons, only a quarter of the strip is discretized. The geometry and the boundary conditions are presented in Fig.~\ref{fig:plaque_geom}. This problem is frequently used in the literature, see for example \cite{BeiraodaVeiga2015, Artioli2017, Auricchio2003, Szabo2009}. The material parameters are a Young modulus  $E = 70~\MPa$, a Poisson ratio $\nu = 0.3$, an initial yield stress $\sigma_{y,0}=0.8~\MPa$, an isotropic hardening modulus $H=10~\MPa$, and a kinematic hardening modulus $K=5~\MPa$. The relative displacement errors $err = ( u- u_{\textrm{ref}} )/ u_{\textrm{ref}}$ versus the number of degrees of freedom are plotted for the points $A$ and $B$ (indicated in Fig.~\ref{fig:plaque_geom}), and for different triangular meshes in Fig.~\ref{fig::plaque_error} (the reference solution $u_{\textrm{ref}}$ is computed with the UPG method and a mesh composed of 37,032 triangles leading to 149,206 degrees of freedom). The relative errors are similar for UPG and HHO(1), and the errors are lower for HHO(2) for the same number of degrees of freedom.  Finally, the equivalent plastic strain $p$ is shown in Fig.~\ref{fig::plaque_p} for UPG and HHO(2) on a triangular mesh. We remark that the results are similar and that there is no sign of locking.
\subsection{Compression of a cube}\label{ss::cube}
\begin{figure}[htbp]
    \centering
    \subfloat[]{
    \label{fig::punch_geom}
        \centering 
        \includegraphics[scale=1.2]{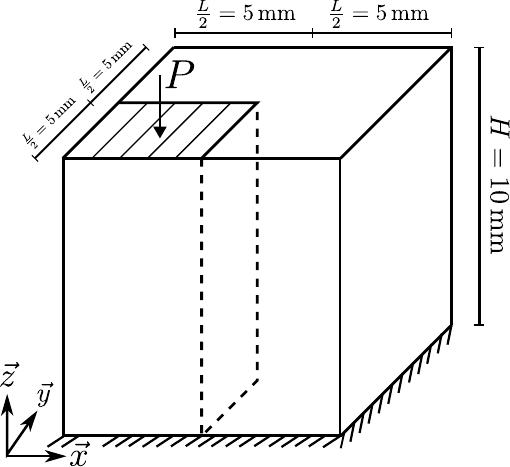} 
  }
    ~ 
    \subfloat[]{
        \centering
	    \includegraphics[scale=0.29]{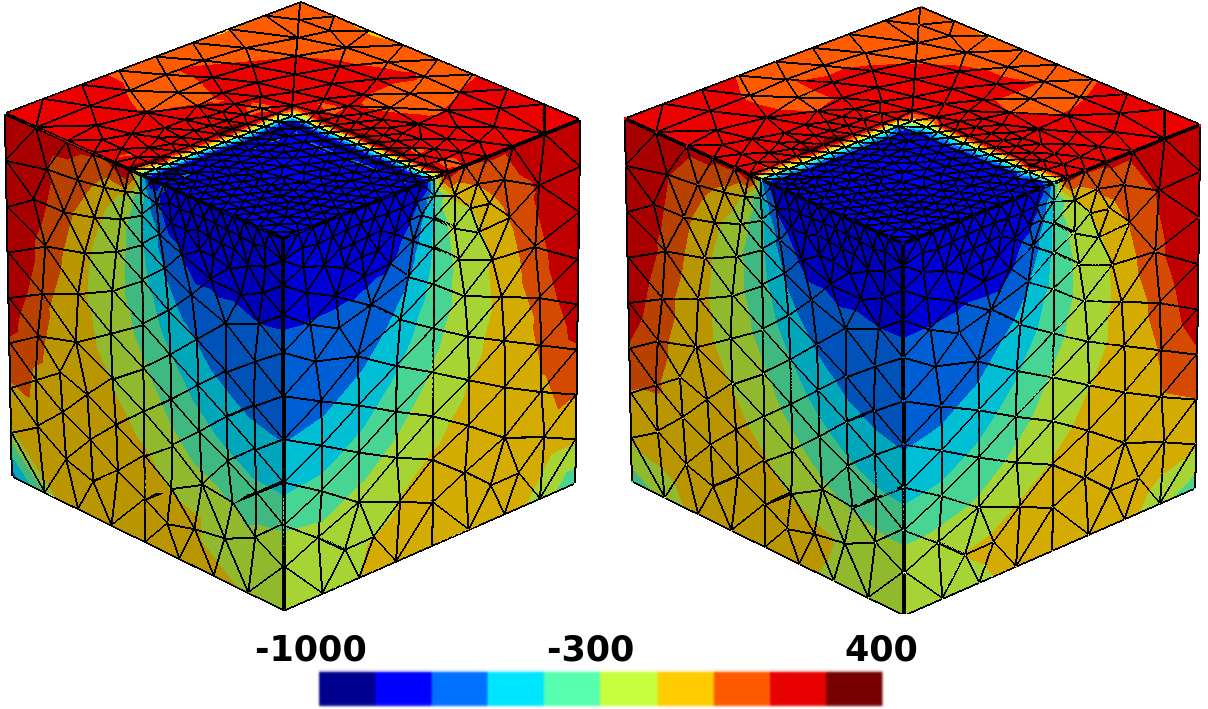}
	    \label{fig::punch_trace}
    }
    \caption{Compression of a cube: (a) Geometry. (b) Trace of the stress tensor on the reference configuration ($\MPa$) with UPG (left) and HHO(1) (right) on a tetrahedral mesh; there are 25,556 degrees of freedom for UPG and 15,947 for HHO(1).}
\end{figure}

This benchmark comes from \cite[Section~5.2]{AlAkhrass2014}. It consists of a rectangular block of length and width $2L= 20~\mm$ and height $H=~10\mm$. The lateral faces are free and the bottom face is clamped. Only one quarter is discretized owing to symmetries, see Fig.~\ref{fig::punch_geom}. The  material parameters are a Young modulus $E = 200~\GPa$, a Poisson ratio $\nu = 0.3$, and an initial yield stress $\sigma_{y,0} = 150~\MPa$. A vertical pressure $P=350~\MPa$ is applied in 30 uniform increments in the part of the upper surface indicated in Fig.~\ref{fig::punch_geom}. The trace of the stress tensor is plotted in Fig.~\ref{fig::punch_trace} for UPG and HHO(1) on a tetrahedral mesh. Both methods do not present oscillations, contrary to the cG formulation (not shown for brevity). 
\subsection{Polyhedral meshes}
\begin{figure}[htbp]
    \centering
    \subfloat[]{
    \label{fig:plaque_mesh}
        \centering 
        \includegraphics[scale=0.29]{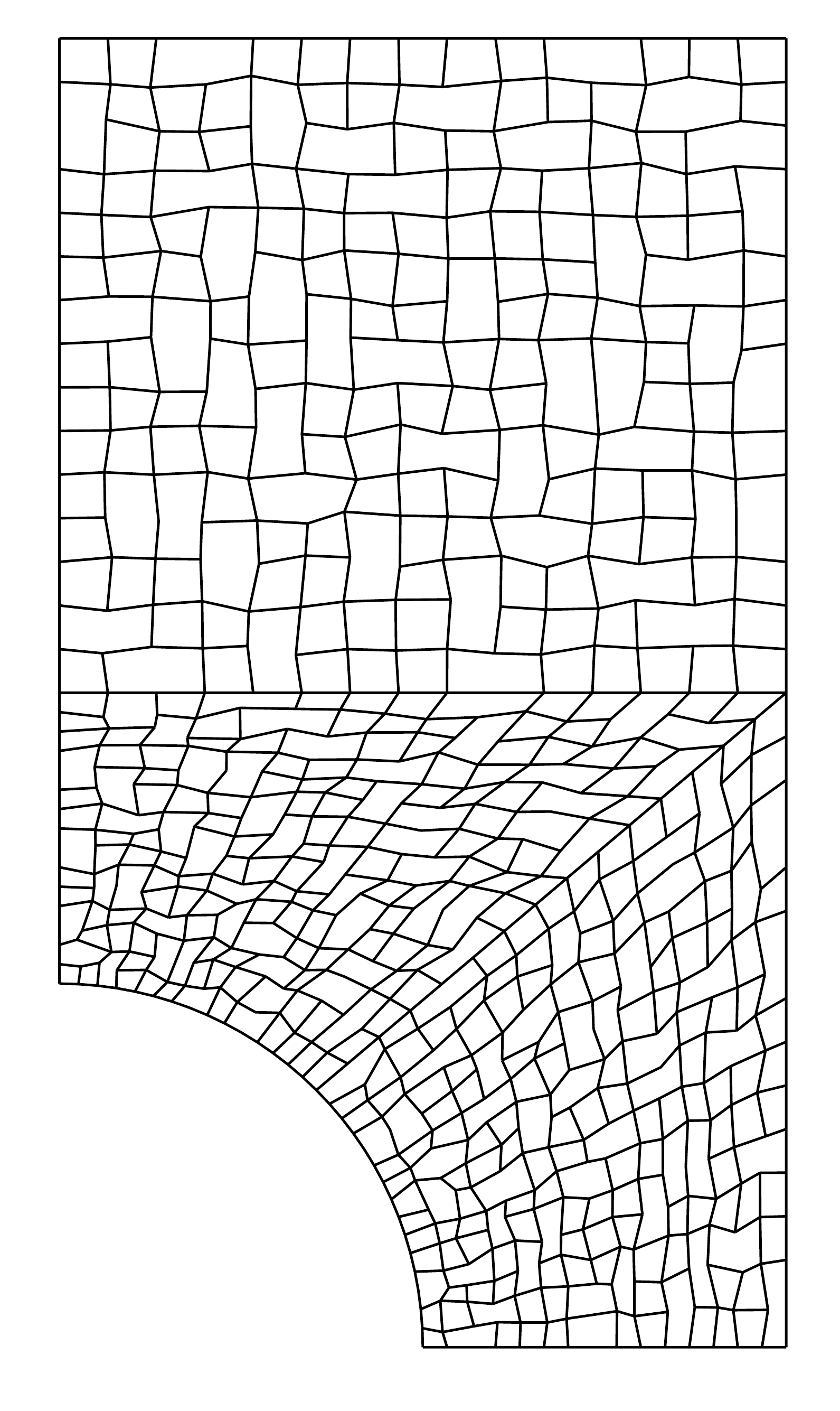} 
  }
    ~ 
    \subfloat[]{
        \centering
	    \raisebox{3mm}{\includegraphics[scale=0.35]{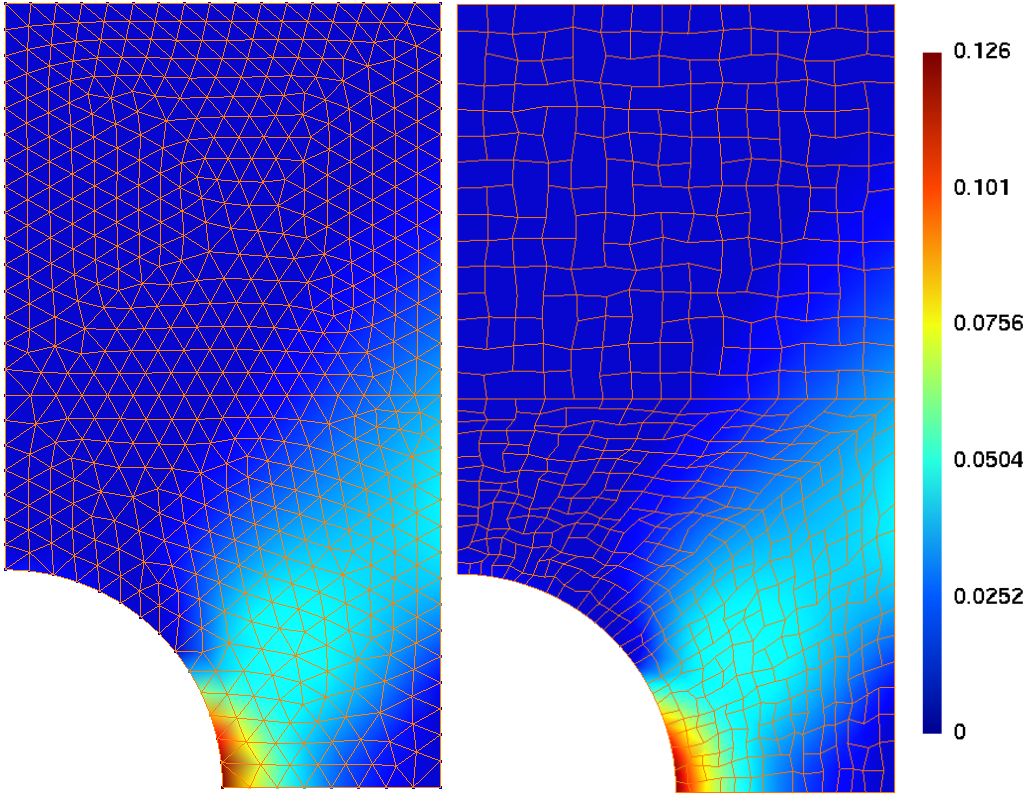}}
	    \label{fig:plaque_p_poly}
    }
    \caption{Perforated strip: (a) Example of a polygonal mesh composed of 536 cells. (b) Equivalent plastic strain $p$ with HHO(2) for a triangular mesh (left) and a polygonal mesh (right); there are 9,750 dofs for the triangular mesh and 7,590 dofs for the polygonal mesh.}
\end{figure}

\begin{figure}[htbp]
    \centering
    \subfloat[]{
    \label{fig::punch_mesh}
        \centering 
        \includegraphics[scale=0.24]{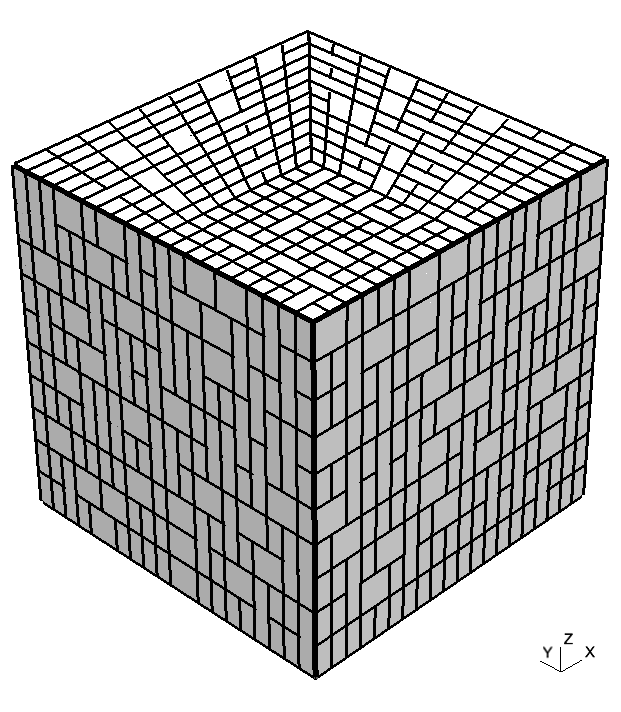} 
  }
    ~ 
    \subfloat[]{
        \centering
	    \includegraphics[scale=0.3]{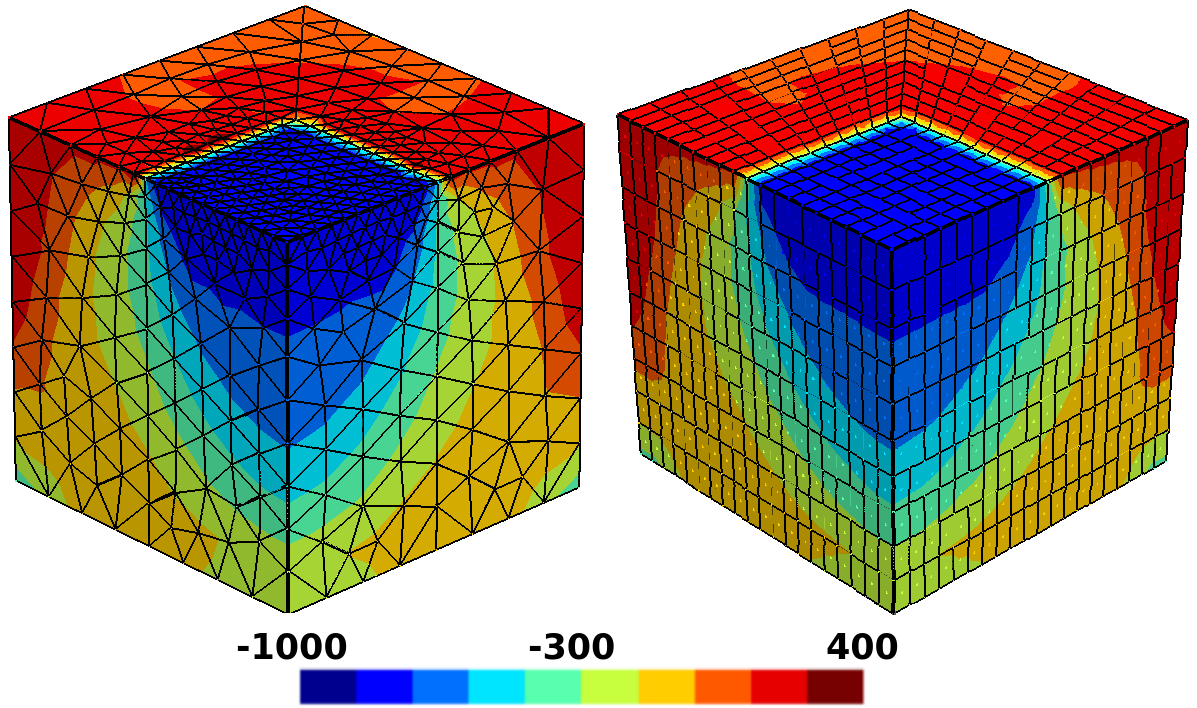}
	    \label{fig::punch_trace_hho1}
    }
    \caption{Compression of a cube:  (a) Example of a polyhedral mesh composed of 2,243 cells. (b) Trace of the stress tensor on the initial configuration (in $\MPa$) for HHO(1) on a tetrahedral mesh (left) and on the polyhedral mesh (right); there are 31,941 dofs for the tetrahedral mesh and 75,261 dofs for the polyhedral mesh.}
\end{figure}

In the previous sections, the proposed HHO method has been tested on simplicial and hexahedral meshes so as to be able to compare it to the UPG method which only supports these types of meshes. Our goal now is to illustrate the fact that the HHO method supports general meshes with possibly non-matching interfaces. For our test cases, the polyhedral meshes are generated from  quadrangular or hexahedral meshes by removing the common face for some pairs of neighbouring cells and then merging the two cells in question (about 30\% of the cells are merged) thereby producing non-matching interfaces materialized by hanging nodes for a significant portion of the mesh cells. In two dimensions, a random moving of the internal nodes is additionally applied in such a way that the cells remain star-shaped with respect to their barycenter. We consider the last two benchmarks presented above, i.e., the perforated strip (see Section~\ref{ss::plaque}) and the compressed cube (see Section~\ref{ss::cube}). Concerning the perforated strip, we show an example of a polyhedral mesh in Fig.~\ref{fig:plaque_mesh} and the equivalent plastic strain for HHO(2) on a triangular mesh and the polygonal mesh in Fig.~\ref{fig:plaque_p_poly}. The results agree very well on the two meshes. 
For the compression of the cube, we use a polyhedral mesh which is generated as above from a hexahedral mesh by removing the common face for some pairs of neighboring cells (see Fig.~\ref{fig::punch_mesh} for the mesh; note that a random moving of the internal nodes is not applied here to avoid nonplanar faces). We compare in Fig.~\ref{fig::punch_trace_hho1} the trace of stress tensor for HHO(1) on a tetrahedral mesh and the polyhedral mesh. The results agree very well on the two meshes, and there is no sign of volumetric locking. These numerical experiments indicate that, as predicted by the theory, the HHO method supports general meshes both in two and three dimensions.
\section{Conclusion}
We have devised and evaluated numerically a Hybrid High-Order method to approximate associative plasticity problems in the small deformation regime. The method shows a robust behavior for the perfect plasticity model as well as for the combined linear isotropic and kinematic hardening model and produces accurate solutions with a moderate number of degrees of freedom. In particular, as mixed methods, the HHO method prevents volumetric locking due to plastic incompressiblity, but with less unknowns than mixed methods for the same accuracy.  Moreover, the HHO method supports general meshes with possibly non-matching interfaces.
This work can be pursued in several directions. One could use a non-local plasticity model, as for example a strain-gradient plasticity model, to take into account scale-dependent effects as in \cite{Djoko2007a}. Furthermore, error estimates can be investigated, possibly by taking inspiration from \cite{Alberty1999,Djoko2007a}, where other discretization methods are analyzed for plasticity problems. Finally, the extension of the present HHO method to elastoplasticity in finite deformations is the subject of ongoing work.
\appendix
\ifCM
\section{Variants of the HHO method}\label{ss:variant}
\else
\section{Appendix}\label{ss:appendix1}
\subsection{Variants of the HHO method}\label{ss:variant}
\fi
Once the polynomial degree $k$ attached to faces has been fixed, there are three different possibilities for the polynomial degree $l$ attached to cells, namely $l \in \{k-1,k,k+1\} \cap \mathbb{N}$ (see \cite{CoDPE:2016} for diffusive problems). Nevertheless, for $k=1$, the choice $l=0$ is not possible for linear elasticity because of the rigid body motions. In this appendix, we focus on the case $k=2$ and we compare the variants $l \in \{1,2,3\}$ for the cell degrees of freedom. We use the notation HHO(2;$l$) and observe that the choice $l=2$ corresponds to the results presented above. For the local operators $\EkT, \DTkp$, and $\SdTk$, the only difference is that we replace $(\vT, \vdT) \in \UkT$ by $(\vT, \vdT) \in \UklT:= \Pld(T;\Reel^d) \times \PkF(\FT; \Reel^d)$.
Moreover, for $l = k + 1$, we consider the simpler expression for $\SdTk: \Poly^{k+1}_{d-1}(\FT, \Reel^d) \rightarrow \PkF(\FT, \Reel^d)$ such that for all  $\vecteur{\theta}\in \Poly^{k+1}_{d-1}(\FT;\Reel^d)$, $\SdTk(\vecteur{\theta}) = \PikdT(\vecteur{\theta})$.
Lemma~\ref{lemma_stability_stab}, Lemma~\ref{lem:equil_sHHO}, and Theorem~\ref{th::coer} remain true (up to minor adaptations). 

\begin{figure}[htbp]
    \centering
    \subfloat[HHO(2,1)]{
        \centering 
        \includegraphics[scale=0.9]{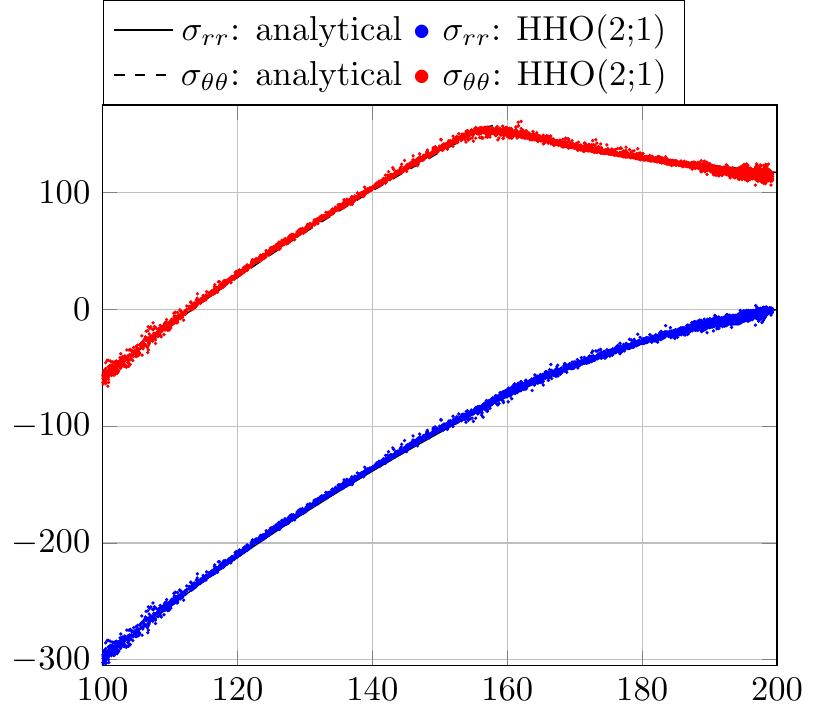} 
  }
    ~ 
    \subfloat[HHO(2,3)]{
        \centering
	    \includegraphics[scale=0.9]{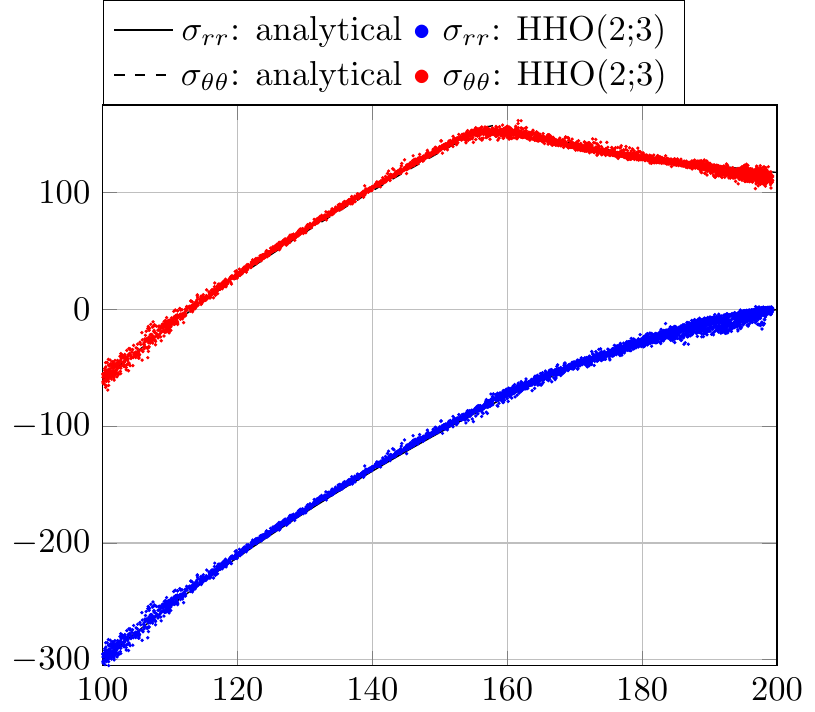}
    }
    \caption{Sphere under internal pressure: $\sigma_{rr}~(\MPa)$ and $\sigma_{\theta \theta}~(\MPa)$ vs. $r~(\mm)$ for HHO(2;1) and HHO(2;3) methods at all the quadrature points for $P=300~\MPa$.}
    \label{fig::efficiency_stress}
\end{figure}

\begin{figure}[htbp]
    \centering
    \subfloat[]{
        \centering 
        \includegraphics[scale=0.9]{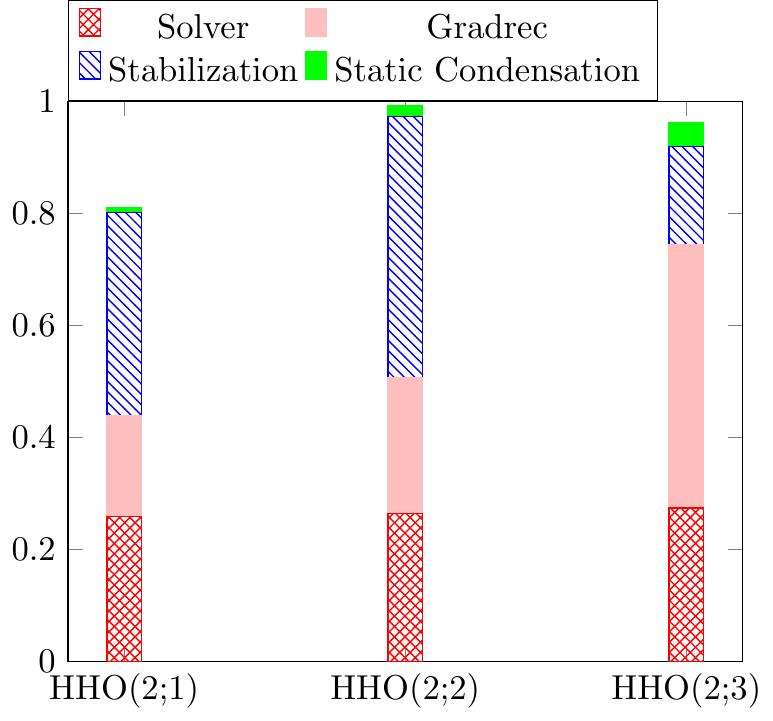} 
        \label{fig:tmp_op}
  }
    ~ 
    \subfloat[]{
        \centering
	    \includegraphics[scale=0.9]{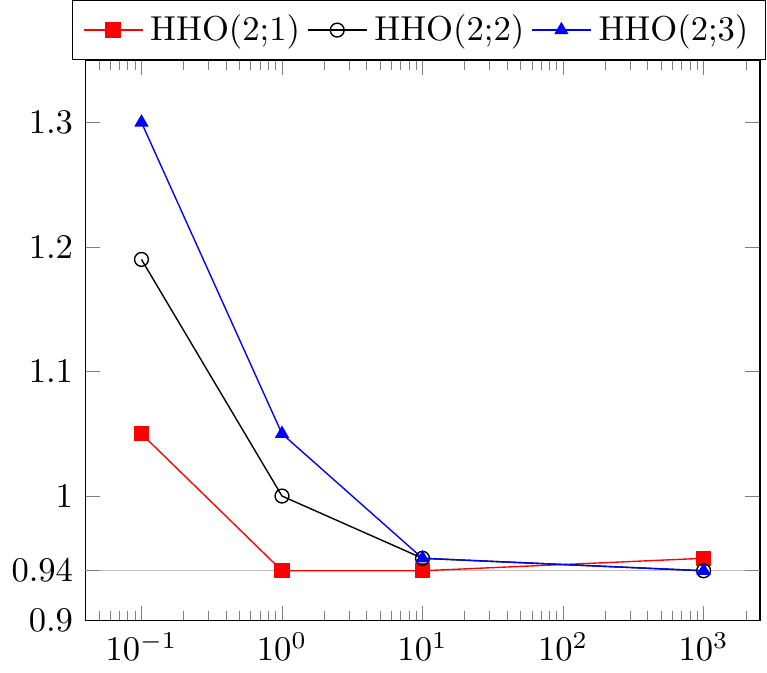} 
	    \label{fig:sphere_beta}
    }
   \caption{Sphere under internal pressure: (a) Comparison of normalized CPU times for  HHO(2;$l$), $l \in \{1,2,3\}$. (b) Total number of  Newton's iterations versus $\beta_0$ normalized by the result for HHO(2;2) and $\beta_0=1$.}
\end{figure}

We compare these HHO variants on the first two benchmarks: the sphere under internal pressure (see Section~\ref{ss::sphere}) and the quasi-incompressible Cook's membrane (see Section~\ref{ss::cook}). For the sphere benchmark, the computed radial ($\sigma_{rr}$) and hoop ($\sigma_{\theta \theta}$) components of the stress tensor are shown in Fig.~\ref{fig::efficiency_stress} at all the quadrature points for $P=300~\MPa$  for HHO(2;1) and HHO(2;3). The results are in agreement with those obtained with HHO(2;2).
Let us now compare the time spent to solve the non-linear problem when using
HHO(2,$l$)  with $l\in\{1,2,3\}$ and $\beta=2\mu$ ($\beta_0=1$).  The assembly time to build the local contributions to the global stiffness matrix  is divided  into three parts: one part, denoted \texttt{Gradrec}, to reconstruct the strain  and build the global system; a second part, denoted \texttt{Stabilization}, to build the stabilization operator (including the time to build the displacement reconstruction, see \eqref{eq_reconstruction_depl}); and a last part, denoted \texttt{Static Condensation}, to perform the static condensation. The solver time, which corresponds to solving the global linear system, is denoted \texttt{Solver}. These times are computed after summation over all the Newton's iterations and are normalized by the total cost associated with HHO(2;2). In Fig.~\ref{fig:tmp_op}, we provide an assessment of the cost on a fixed mesh with 506 tetrahedra. We observe that the difference between HHO(2;2) and HHO(2;3) is not really important; in fact, the time that HHO(2;3) spends to reconstruct the less expensive stabilization is compensated by a larger number of Newton's iterations.  The HHO(2;1) variant turns out to be the most efficient (around 20\% less CPU time  than HHO(2;2)); indeed it needs less Newton's iterations and the cost of a single Newton's iteration is the cheapest.  In Fig.~\ref{fig:sphere_beta}, we report the total number of Newton's iterations normalized by the result for HHO(2;2) and $\beta_0=1$ versus $\beta_0$. On the one hand, we remark that $\beta_0$ has a significant influence on the number of Newton's iterations if $\beta_0 \lesssim 1$ and on the other hand, the different variants need the same number of Newton's iterations to converge if $\beta_0 \gtrsim 10$. Note that this experiment is particularly challenging since  we are considering here perfect plasticity for which the stability result from Theorem~\ref{th::coer} is not applicable.

\begin{figure}[htbp]
    \centering
    \subfloat[]{
        \centering 
        \includegraphics[scale=0.9]{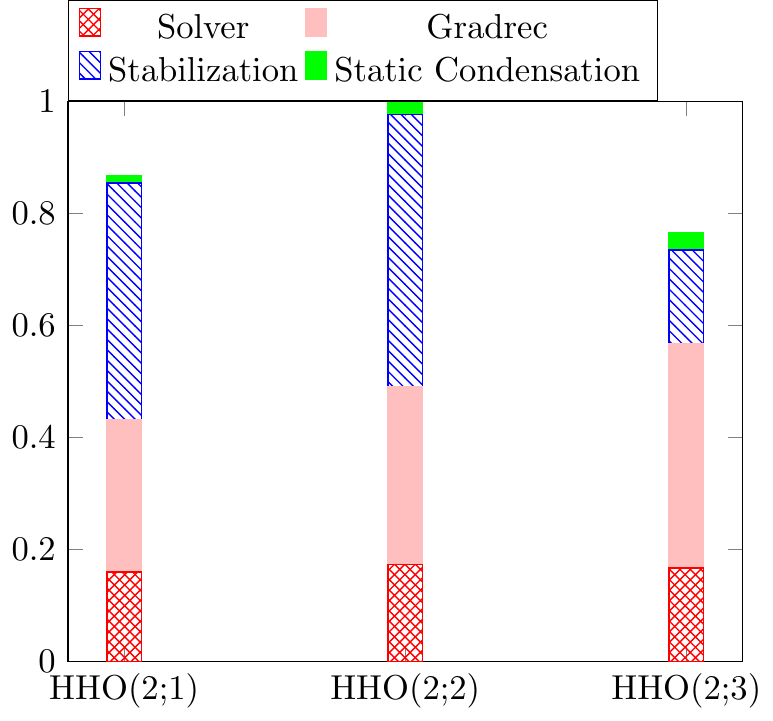} 
        \label{fig:tmp_cook_op}
  }
    ~ 
    \subfloat[]{
        \centering
	    \includegraphics[scale=0.9]{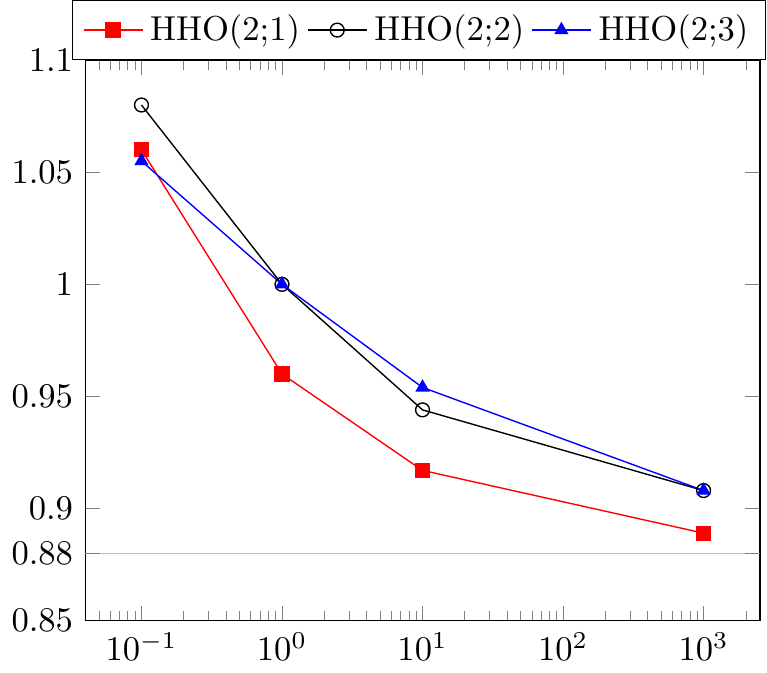} 
	    \label{fig:cook_beta}
    }
   \caption{Quasi-incompressible Cook's membrane: (a) Comparison of normalized CPU times for  HHO(2;$l$), $l \in \{1,2,3\}$. (b) Number of Newton's iteration versus $\beta_0$ normalized by the result for HHO(2;2) and $\beta_0=1$.}
\end{figure}

We repeat the above experiments with the Cook's membrane problem with strain hardening plasticity so that the stability result of Theorem~\ref{th::coer} holds true with $\theta_{\Th, Q}\approx 0.25$. In Fig.~\ref{fig:tmp_cook_op}, the CPU times on a $64 \times 64$ quadrangular mesh are reported. Here, HHO(2;3) turns out to be the most efficient variant (around  22\% less CPU time than HHO(2,2)); indeed it needs the same number of Newton's iterations and the computation of the stabilization operator is faster than for the other variants. Nevertheless, these differences in terms of total CPU times are less  noticeable in two dimensions than in three dimensions since the computations are less intensive. In Fig.~\ref{fig:cook_beta}, we plot the number of Newton's iterations normalized by the result for HHO(2;2) and $\beta_0=1$ versus $\beta_0$.  As above, HHO(2;1)  is the variant that depends the least on $\beta_0$ but this behavior is less pronounced than for the sphere benchmark. In addition, whatever variant is used, if we take $\beta_0 =  \frac{\theta_{\Th, Q}}{2\mu} \approx 0.005$, the Newton's method needs more than 500 iterations to converge (compared to 108 iterations for HHO(2;2) and $\beta_0=1$). A first conclusion is that it seems reasonable to take $\beta_0 \in [1, 100]$ since the  number of Newton's iterations  is lower and the condition number does not increase too much. A second conclusion is that it seems preferable to use HHO(2;1) than HHO(2;2) or HHO(2;3) since HHO(2;1) is less sensitive to the choice of $\beta_0$. 
\ifCM
\section{Proof of Lemma~\ref{lem:equil_sHHO}}\label{proof_dw}
\else
\subsection{Proof of Lemma~\ref{lem:equil_sHHO}}\label{proof_dw}
\fi

\begin{Proof}
Recall the notation $\dstress_T^n =  (\dstress_{T}^n(\Qp_{T,j}))_{1 \leq j \leq m_Q}$ with $\dstress_{T}^n(\Qp_{T,j}) = \elasticmodule : (\EkT(\uTn,\udTn)(\Qp_{T,j}) - \tilde{\strain}^{p,n}_T(\Qp_{T,j}))$ for all $1 \leq j \leq m_Q$.
Let us consider the virtual displacement $((\dvT \delta_{T,T'})_{T'\in\Th},(\vecteur{0})_{F\in\Fh})\in\Ukhz$ in~\eqref{discrete_problem_ptv}, with the Kronecker delta such that $\delta_{T,T'}=1$ if $T=T'$ and $\delta_{T,T'}=0$ otherwise. Owing to~\eqref{eq_reconstruction_grad}, and since the quadrature is by assumption at least of order $2k$, we have
\begin{align*}
\psv[T]{ \loadext^n}{\dvT} 
 ={} & 
\psmd[T]{\dstress^n_T}{\EkT(\dvT,\vecteur{0})} +   \beta \psv[\dT]{\gamma_{\dT}\SdTk(\udTn - \uTn{}_{|\dT})}{\SdTk(-\dvT{}_{|\dT})} \\
 ={}& \psm[T]{\PikTmd (\dstress^n_T)}{\EkT(\dvT,\vecteur{0})} - \beta \psv[\dT]{\SdTks(\gamma_{\dT}\SdTk(\udTn -\uTn{}_{|\dT}))}{\dvT{}_{|\dT}} \\
={} & \psm[T]{\PikTmd (\dstress^n_T)}{\grads{\dvT}} - \psv[\dT]{\PikTmd( \dstress^n_T) \, \SCAL \nT}{\dvT} \\
& -  \beta\psv[\dT]{\SdTks(\gamma_{\dT}\SdTk(\udTn-\uTn{}_{|\dT}))}{\dvT{}_{|\dT}}\\
={}&\psmd[T]{\dstress^n_T}{\grads{\dvT}} -
\psv[\dT]{\vecteur{T}_{T}^n}{\dvT}.
\end{align*}
This establishes the local principle of virtual work~\eqref{eq:pwk}.
Similarly, the law of action and reaction~\eqref{eq:balance} follows by considering, for all $F\in\Fhi\cup\Fhbn$, the virtual displacement $((\vecteur{0})_{T\in\Th},(\dvF \delta_{F,F'})_{F'\in\Fh})\in\Ukhz$ in~\eqref{discrete_problem_ptv} (with obvious notation for the face-based Kronecker delta), and observing that both $\dvF$ and $\vecteur{T}_{T_{\pm}|F}$ are in $\PkF(F;\Reel^d)$. If $F \in \Fhbn$ with $F=\dT\cap \Bn\cap H_F$, we have
\begin{align*}
\psv[F]{ \Tn^n}{\dvF} 
={} &\psv[F]{ \PikF (\Tn^n)}{\dvF} \\
 ={} & \psmd[T]{\dstress^n_T}{\EkT(\vecteur{0}, \dvF)} +   \beta \psv[F]{\gamma_{\dT}\SdTk(\udTn - \uTn{}_{|\dT})}{\SdTk(\dvF)} \\
 ={}& \psm[T]{\PikTmd (\dstress^n_T)}{\EkT(\vecteur{0}, \dvF)} +     \beta \psv[\dT]{\gamma_{\dT}\SdTk(\udTn - \uTn{}_{|\dT})}{\SdTk(\dvF)} \\
={} &  \psv[F]{\PikTmd (\dstress^n_T) \, \SCAL \nT}{\dvF} +  \beta \psv[\dT]{\SdTks(\gamma_{\dT}\SdTk(\udTn-\uTn{}_{|\dT}))}{\dvF}\\
={}& \psv[F]{\vecteur{T}_{T|F}^n}{\dvF},
\end{align*}
whereas if $F \in \Fhi$ with $F=\partial T_- \cap \partial T_+ \cap H_F$, we have
\begin{align*}
0 = {} & \psmd[T_{+}]{\dstress^n_{T_{+}}}{\EkTp(\vecteur{0}, \dvF)} +   \beta \psv[\partial T_+]{\gamma_{\dT_+}\SdTpk(\udTpn - \uTpn{}_{|\dT_+})}{\SdTpk(\dvF)} \\
 &+ \psmd[T_{-}]{\dstress^n_{T_{-}}}{\EkTm(\vecteur{0}, \dvF)} +   \beta \psv[\partial T_-]{\gamma_{\dT_-}\SdTmk(\udTmn - \uTmn{}_{|\dT_-})}{\SdTmk(\dvF)} \\
 ={} &  \psv[\partial T_+]{\PikTpmd (\dstress^n_{T_{+}}) \, \SCAL \nTp}{\dvF} +  \beta \psv[\partial T_+]{\SdTpks(\gamma_{\dT_+}\SdTpk(\udTpn-\uTpn{}_{|\dT_+}))}{\dvF}\\
&+  \psv[\partial T_-]{\PikTmmd (\dstress^n_{T_{-}}) \, \SCAL \nTm}{\dvF}  +  \beta \psv[\partial T_-]{\SdTmks(\gamma_{\dT_-}\SdTmk(\udTmn-\uTmn{}_{|\dT_-}))}{\dvF} \\
= {}& \psv[F]{\vecteur{T}_{{T_{+}}|F}^n + \vecteur{T}_{{T_{-}}|F}^n}{\dvF}.
\end{align*}
\end{Proof}
\ifCM
\section{Proof of Theorem~\ref{th::coer}}\label{proof_coer}
\else
\subsection{Proof of Theorem~\ref{th::coer}}\label{proof_coer}
\fi
\begin{Proof}
Since the material is strain-hardening, the consistent elastoplastic tangent modulus  is symmetric positive-definite (see line~\ref{line_Cep} of Algorithm~\ref{algo::plasticity}). Hence, its smallest eigenvalue is real and positive, so that $\theta_{\Th, Q} >0$. Observing that $\EkT ( \vT, \vdT) \in \Pkd(T, \Reel^{d \times d})$ for all $\vh\in \Ukhz$ and that all the quadrature weights are by assumption positive, we infer that
\begin{align*}
&  \sum_{T\in\Th} \psmd[T]{\depmodule: \EkT(\vT,\vdT)}{\EkT(\vT,\vdT)} \\ 
&  \hphantom{+} +  \sum_{T\in\Th} \beta \psv[\dT]{\gamma_{\dT}\SdTk(\vdT- \vT{}_{|\dT})}{\SdTk(\vdT -\vT{}_{|\dT})} \\
& =  \sum_{T\in\Th} \sum_{j=1 }^{m_Q} \Wp_{T,j}  \EkT(\vT,\vdT)(\Qp_{T,j}) : \depmodule(\Qp_{T,j}) : \EkT(\vT,\vdT)(\Qp_{T,j})  \\ 
&  \hphantom{+}+  \sum_{T\in\Th} \beta \psv[\dT]{\gamma_{\dT}\SdTk(\vdT- \vT{}_{|\dT})}{\SdTk(\vdT -\vT{}_{|\dT})} \\
& \geq  \sum_{T\in\Th} \left\lbrace\theta_{\Th, Q}  \sum_{j=1 }^{m_Q} \Wp_{T,j} \EkT(\vT,\vdT)(\Qp_{T,j}) : \EkT(\vT,\vdT)(\Qp_{T,j})  \right\rbrace \\ 
&  \hphantom{+}+  \sum_{T\in\Th} \beta \psv[\dT]{\gamma_{\dT}\SdTk(\vdT- \vT{}_{|\dT})}{\SdTk(\vdT - \vT{}_{|\dT})}\\
&\geq \min(\theta_{\Th, Q}, \beta) \sum_{T\in\Th}  \left\lbrace  \normemd[T]{\EkT (\vT,\vdT)}^2 +    \normev[\dT]{\gamma_{\dT}^{\frac12}\SdTk(\vdT - \vT{}_{|\dT})}^2 \right\rbrace \\
&= \min(\theta_{\Th, Q} , \beta) \sum_{T\in\Th}  \left\lbrace  \normem[T]{\EkT (\vT,\vdT)}^2 +   \normev[\dT]{\gamma_{\dT}^{\frac12}\SdTk(\vdT - \vT{}_{|\dT})}^2 \right\rbrace.
\end{align*}
We conclude by using the stability result from Lemma~\ref{lemma_stability_stab} and recalling that $\beta=2\mu\beta_0$.
\end{Proof}
\ifCM
\bibliographystyle{elsarticle-num}{\noindent \bf{References}}
\else
\bibliographystyle{abbrvnat}
\fi
\bibliography{Bibliographie}
\end{document}